\newcommand{\bea}{\begin{eqnarray}}
\newcommand{\eea}{\end{eqnarray}}
\newcommand{\be}{\begin{equation}}
\newcommand{\ee}{\end{equation}}
\newcommand{\ben}{\begin{enumerate}}
\newcommand{\een}{\end{enumerate}}
\newcommand{\bi}{\begin{itemize}}
\newcommand{\ei}{\end{itemize}}
\newcommand{\bmi}[1]{\begin{minipage}{#1 cm}}
\newcommand{\emi}{\end{minipage}}

\def\eck#1{\left\lbrack #1 \right\rbrack}

\def\rund#1{\left( #1 \right)}
\def\abs#1{\left\vert #1 \right\vert}

\def\ave#1{\left\langle #1 \right\rangle}

\def\Re{{\cal R}\hbox{e}}
\def\Im{{\cal I}\hbox{m}}
\def\A{{\cal A}}
\def\B{{\cal B}}

\def\R{{\cal R}}

\def\d{{\rm d}}

\def\hb{{\hfill\break}}

\def\vp{\varphi}
\def\vt{{\vartheta}}

\def\Real{{\rm I\mathchoice{\kern-0.70mm}{\kern-0.70mm}{\kern-0.65mm}%
  {\kern-0.50mm}R}}
\def\C{\rm C\kern-.42em\vrule width.03em height.58em depth-.02em
       \kern.4em}
\font \bolditalics = cmmib10
\def\bx#1{\leavevmode\thinspace\hbox{\vrule\vtop{\vbox{\hrule\kern1pt
        \hbox{\vphantom{\tt/}\thinspace{\bf#1}\thinspace}}
      \kern1pt\hrule}\vrule}\thinspace}

\def \vc #1{{\textfont1=\bolditalics \hbox{$\bf#1$}}}
{\catcode`\@=11
\gdef\SchlangeUnter#1#2{\lower2pt\vbox{\baselineskip 0pt \lineskip0pt
  \ialign{$\m@th#1\hfil##\hfil$\crcr#2\crcr\sim\crcr}}}
}

\def\lesssim{\mathrel{\mathpalette\SchlangeUnter<}}

\def\ueber#1#2{{\setbox0=\hbox{$#1$}%
  \setbox1=\hbox to\wd0{\hss$\scriptscriptstyle #2$\hss}%
  \offinterlineskip
  \vbox{\box1\kern0.4mm\box0}}{}}

\def\bx#1{\leavevmode\thinspace\hbox{\vrule\vtop{\vbox{\hrule\kern1pt
        \hbox{\vphantom{\tt/}\thinspace{\bf#1}\thinspace}}
      \kern1pt\hrule}\vrule}\thinspace}

\def\llabel#1{\label{sc:#1}}
\def\elabel#1{\label{eq:#1}}

{\catcode`\@=11
\gdef\SchlangeUnter#1#2{\lower2pt\vbox{\baselineskip 0pt \lineskip0pt
  \ialign{$\m@th#1\hfil##\hfil$\crcr#2\crcr\sim\crcr}}}
}

\def\lesssim{\mathrel{\mathpalette\SchlangeUnter<}}

\documentclass[onecolumn]{aa}
\usepackage{graphicx,rotating}
\usepackage{pstricks,pst-node}
\begin{document}
   \title{The three-point correlation function of cosmic shear. II:
   Relation to the bispectrum of the projected mass density and generalized
   third-order aperture measures}

   \author{
  Peter Schneider
          \inst{1}
          \and
          Martin Kilbinger\inst{1} 
          \and
          Marco Lombardi\inst{1,2}
          }

   \offprints{P. Schneider}

   \institute{Institut f. Astrophysik u. Extr. Forschung, Universit\"at Bonn,
              Auf dem H\"ugel 71, D-53121 Bonn, Germany\\
              \email{peter@astro.uni-bonn.de}
         \and
              European Southern Observatory, Karl-Schwarzschild-Str.\ 2,
              D-85741 Garching, Germany
             }
\titlerunning{The three-point correlation function of cosmic shear II}
\authorrunning{P. Schneider, M. Kilbinger \& M. Lombardi}

   \date{Received ; accepted }

   \abstract{
        Cosmic shear, the distortion of images of high-redshift sources by the
        intervening inhomogeneous matter distribution in the Universe, has
        become one of the essential tools for observational cosmology since
        it was first measured in 2000. Since then, several surveys have been
        conducted and analyzed in terms of second-order shear
        statistics. Current surveys are on the verge of providing useful
        measurements of third-order shear statistics, and ongoing and future
        surveys will provide accurate measurements of the shear three-point
        correlation function which contains essential information about the
        non-Gaussian properties of the cosmic matter distribution.

        We study the relation of the three-point cosmic shear statistics to
        the third-order statistical properties of the underlying convergence,
        expressed in terms of its bispectrum. Explicit relations for the
        natural components of the shear three-point correlation function
        (which we defined in an earlier paper) in terms of the bispectrum are
        derived. The behavior of the correlation function under parity
        transformation is obtained and found to agree with previous
        results. We find that in contrast to the two-point shear correlation
        function, the three-point function at a given angular scale $\theta$
        is not affected by power in the bispectrum on much larger scales. These
        relations are then inverted to obtain the bispectrum in terms of the
        three-point shear correlator; two different expressions, corresponding
        to different natural components of the shear correlator, are obtained
        and can be used to separate E and B-mode shear contributions. These
        relations allow us to explicitly show that correlations containing an
        odd power of B-mode shear vanish for parity-symmetric
        fields. Generalizing a recent result by Jarvis et al., we derive
        expressions for the third-order aperture measures, employing multiple
        angular scales, in terms of the (natural components of the)
        three-point shear correlator and show that they contain essentially
        all the information about the underlying bispectrum. We discuss the
        many useful features these (generalized) aperture measures have that
        make them convenient for future analyses of the skewness of the cosmic
        shear field (and any other polar field, such as the polarization of
        the Cosmic Microwave Background).

   \keywords{cosmology -- gravitational lensing -- large-scale
                structure of the Universe
               }
   }

   \maketitle
%

\section{\llabel{intro}Introduction}
Recent surveys have measured second-order cosmic shear statistics with
high accuracy, owing to the large sky area covered, and thus the large
number of faint galaxy images (e.g., van Waerbeke et al.\ 2001, 2002;
Jarvis et al.\ 2003a; Hoekstra et al.\ 2002). With surveys of this
size, it now becomes feasible to obtain higher-order cosmic shear
statistics which probe the non-Gaussian features of the cosmic shear
field. These higher-order statistics are particularly useful in
breaking near-degeneracies in cosmological parameters which are
present at the level of second-order statistics. Bernardeau et al.\
(1997), van Waerbeke et al.\ (1999) and others pointed out that the
skewness of the convergence underlying the cosmic shear field can
break the degeneracy between the density parameter $\Omega_{\rm m}$
and the normalisation of the matter power spectrum, expressed in terms
of the r.m.s. density fluctuations $\sigma_8$ on a scale of $8
h^{-1}\,{\rm Mpc}$. However, the convergence cannot be observed
directly, but needs to be inferred from the observed galaxy image
ellipticities which yield an estimate of the local shear. The
dispersion of the shear in a (circular) aperture, frequently used to
quantify second-order shear statistics, cannot be generalized to
a third-order statistics. Schneider
et al.\ (1998; hereafter SvWJK) have defined an alternative cosmic
shear measure, the aperture mass, which is a scalar quantity that can
be directly obtained from the shear, and therefore is particularly
suited to define higher-order statistics.

Recently, interest in higher-order cosmic shear statistics has been
revived. The three-point correlation function (3PCF henceforth) of the shear
contains all the information on the third-order statistical properties of the
shear field, and therefore is of prime interest. In addition, it can be
obtained directly from the observed image ellipticities and, in contrast to
the aperture mass statistics, is insensitive to holes and gaps in the data
field. However, the shear 3PCF is a function with $2\times 2\times 2=8$
components (since each shear has two independent components) and 3
variables (e.g., the sides of the triangle formed by the three points)
and therefore difficult to handle. Bernardeau et al.\ (2003) defined a
specific integral over the 3PCF and applied that to the VIRMOS-DESCART survey
in Bernardeau et al.\ (2002) to obtain the first detection of a non-zero
third-order cosmic shear signal. Using the same observational data, Pen et
al.\ (2003) calculated the skewness of the aperture mass, where the latter has
been obtained from integrating the shear 3PCF. Jarvis et al.\ (2003b, JBJ
hereafter) obtained an alternative expression for the aperture mass skewness
in terms of the shear 3PCF and applied this to the CTIO cosmic shear survey,
finding a signal at about the 2.3-$\sigma$ level.

Following a different approach, the 3PCF was considered directly in a
number of recent papers. Schneider \& Lombardi (2003; hereafter Paper\
I) defined special combinations of the shear 3PCF which we termed
the `natural components', because they obey simple
transformation laws under coordinate rotations. In particular, we
derived the behavior of the 3PCF under parity transformations, and
showed that all eight components are expected to be non-zero for a
general triangle configuration. Zaldarriaga \& Scoccimarro (2003) and
Takada \& Jain (2003a) obtained analytic approximations and numerical
results, using ray-tracing simulations, for the 3PCF.  Takada \& Jain
(2003b) provide an extensive study of the expectation for the shear 3PCF
in terms of the halo model of the large-scale structure, and they
verified the accuracy of their analytical results with numerical
simulations. Schneider (2003) investigated the transformation
properties of a general 3PCF of a polar under parity transformations
and showed that the expectation value of any quantity containing an
odd power of B-modes vanishes for parity-invariant shear fields.

In this paper, we first consider the relation between the shear 3PCF and the
bispectrum of the underlying convergence (or projected density)
field. If the shear field is derivable from a scalar potential (that
is, a pure E-mode field), as expected for cosmic shear in the absence
of intrinsic galaxy alignments and systematics in the observing
process, the bispectrum of the convergence fully encodes the
third-order information of the random field.\footnote{It should be
noted that even for a pure cosmic shear field, B-modes do occur if
source galaxies are clustered (Bernardeau 1998; Schneider et al.\
2002); furthermore, they can occur from slight violations of the lowest-order
approximations employed when considering light propagation through a
slightly inhomogeneous Universe -- see, e.g., Bernardeau et al.\ (1997),
SvWJK, Jain et al.\ (2000).} In this first part of the paper, we
therefore generalize the relations between the power spectrum of the
convergence and the various second-order shear statistics (derived in
Crittenden et al.\ 2002; Schneider et al.\ 2002) to third-order shear
statistics. After some
preliminaries in Sect.~2, we derive the shear 3PCF in terms of the
bispectrum of the convergence. From these explicit relations, general
transformation laws of the 3PCF can be directly seen; for example, the
behavior of the 3PCF under parity inversion as studied before in Paper~I and
in Schneider (2003) can be explicitly
verified, as will be shown in Sect.~4. In Sect.~5 we invert these
relations, i.e., we express the bispectrum in terms of the 3PCF of the
shear. We obtain two formally different expressions for the bispectrum
which must, 
however, be identical in the case of a pure E-mode shear field.

In the second part of this paper (Sect.~6), we consider the
third-order aperture mass statistics as a particularly convenient
integral over the shear 3PCF; in fact, this part of the paper will
quite likely be most relevant for future studies of higher-order
cosmic shear statistics.  We first express $\ave{M_{\rm
ap}^3(\theta)}$ in terms of the bispectrum and then replace 
the bispectrum in terms of the 3PCF. This procedure
yields the same result for $\ave{M_{\rm ap}^3(\theta)}$ in terms of
the shear 3PCF as derived by JBJ. We then argue that the third-order
aperture measures contain only part of the information about the
bispectrum of the underlying convergence, and generalize the aperture
measures to the case of three different scale lengths. We show that
this generalization allows us to obtain essentially the full
information about the bispectrum. These generalized aperture measures
are then expressed in terms of the shear 3PCF. Sect.~\ref{sc:summary}
summarizes and discusses our results.

It must be stressed that all our results are valid for other random
fields which share the properties of that of cosmic shear: A
homogeneous, isotropic, parity-symmetric random field of a polar. The
most obvious example of such a field in the cosmological context,
apart from cosmic shear, is the polarization field of the Cosmic
Microwave Background (e.g., Zaldarriaga et al.\ 1997)

\section{\llabel{preli}Preliminaries}
In the first part of this paper (through Sect.~\ref{sc:Bispect}) we shall
consider a shear field $\gamma$ which is caused by an underlying projected
density (or convergence) field $\kappa$, as is expected for a shear field
produced by light propagation in an inhomogeneous Universe (e.g., Blandford et
al.\ 1991; Miralda-Escud\'e 1991; Kaiser 1992; see also the reviews by
Mellier 1999 and Bartelmann \& Schneider 2001). The relation between the shear
(expressed throughout this paper as a complex number) and the convergence is
most simply given in Fourier space,
\be
\hat\gamma(\vc\ell)={\rm e}^{2{\rm i}\beta}\,\hat \kappa(\vc \ell)\;,
\elabel{FTrel}
\ee
where $\beta$ is the polar angle of $\vc\ell$, and $\vc\ell$ is the Fourier
transform variable of the angular position vector on the sky. 

In Paper I we considered the 3PCF of the shear. Since the shear is a
two-component quantity, the 3PCF has 8 independent components. Since one
cannot form a scalar from the product of three shears alone, one needs to
project the shear with respect to some reference directions. The three points
at which the shear is considered form a triangle, and one can project the
shear along directions attached to such a triangle, i.e., which rotate with
the triangle in coordinate rotations. We have considered a number of such
obvious projections, namely with respect to the directions of the vertices
towards one of the centers of a triangle. Let $\zeta_i$ be the polar angle of
the vector connecting the point $\vc X_i$ with the chosen center, then the
Cartesian components $\gamma_{1,2}$ of the shear $\gamma=\gamma_1+{\rm
i}\gamma_2$ are used to define the tangential and cross components of the
shear relative to the chosen direction $\zeta_i$,
\be
\gamma(\vc X_i;\zeta_i)\equiv
\gamma_{\rm t}(\vc X_i;\zeta_i)+{\rm i} \gamma_\times(\vc X_i;\zeta_i)
=-\eck{\gamma_1(\vc X_i)+{\rm i}\gamma_2(\vc X_i)}\,{\rm e}^{-2{\rm
i}\zeta_i}
\equiv
-\gamma(\vc X_i)\,{\rm e}^{-2{\rm
i}\zeta_i} \;.
\ee
If the reference directions are changed, the tangential and cross
components of the 
shear will change correspondingly. In particular, defining the 3PCF of the
shear components, they will change if a different center of the triangle is
chosen. In Paper~I we defined four complex `natural components' of the
shear 3PCF which show a simple behavior under such transformations; they have
been termed $\Gamma^{(\mu)}(x_1,x_2,x_3)$, $\mu=0,1,2,3$. The $\Gamma^{(i)}$,
$i=1,2,3$ can be obtained from one another by permutations of the arguments
$x_i$, which represent the sides of the triangle.

\begin{figure}[!t]
  \parbox[t]{0.59\hsize}{%
    \resizebox{\hsize}{!}{\SpecialCoor
\def\Ax{1 }\def\Ay{0 }%
\def\Bx{9 }\def\By{2 }%
\def\Cx{2 }\def\Cy{6 }%
\def\a{\Bx \Cx sub dup mul \By \Cy sub dup mul add sqrt }%
\def\b{\Cx \Ax sub dup mul \Cy \Ay sub dup mul add sqrt }%
\def\c{\Ax \Bx sub dup mul \Ay \By sub dup mul add sqrt }%
\def\A{\Cy \Ay sub \Cx \Ax sub atan \By \Ay sub \Bx \Ax sub atan sub }%
\def\B{\Ay \By sub \Ax \Bx sub atan \Cy \By sub \Cx \Bx sub atan sub }%
\def\C{\By \Cy sub \Bx \Cx sub atan \Ay \Cy sub \Ax \Cx sub atan sub }%
\def\r{\b \c add \a sub \c \a add \b sub \a \b add \c sub mul mul
\a \b \c add add div sqrt 2 div }%
\def\R{\a \A sin div 2 div }%
\def\Area{\a \b \C sin mul mul 2 div }
\def\ha{\Area 2 mul \a div }
\def\hb{\Area 2 mul \b div }
\def\hc{\Area 2 mul \c div }
\def\HAx{\Ax \ha \Hy \Ay sub \Hx \Ax sub atan cos mul add }
\def\HAy{\Ay \ha \Hy \Ay sub \Hx \Ax sub atan sin mul add }
\def\HBx{\Bx \hb \Hy \By sub \Hx \Bx sub atan cos mul add }
\def\HBy{\By \hb \Hy \By sub \Hx \Bx sub atan sin mul add }
\def\HCx{\Cx \hc \Hy \Cy sub \Hx \Cx sub atan cos mul add }
\def\HCy{\Cy \hc \Hy \Cy sub \Hx \Cx sub atan sin mul add }
\def\baricentricX#1#2#3{\Ax #1 mul \Bx #2 mul \Cx #3 mul add add #1 #2
  #3 add add div }%
\def\baricentricY#1#2#3{\Ay #1 mul \By #2 mul \Cy #3 mul add add #1 #2
  #3 add add div }%
\def\trilinearX#1#2#3{\baricentricX{#1 \a mul }{#2 \b mul }{#3 \c mul }}%
\def\trilinearY#1#2#3{\baricentricY{#1 \a mul }{#2 \b mul }{#3 \c mul }}%
\def\Hx{\trilinearX{\B cos \C cos mul }{\C cos \A cos mul }{\A cos \B cos mul }}%
\def\Hy{\trilinearY{\B cos \C cos mul }{\C cos \A cos mul }{\A cos \B cos mul }}%
\psset{unit=0.8cm}%
\begin{pspicture}(0,-0.2)(11.5,7.2)
  \pnode(! \Ax \Ay){X_1}
  \pnode(! \Bx \By){X_2}
  \pnode(! \Cx \Cy){X_3}
  \pnode(! \Hx \Hy){H}
  \pnode(! \Bx \Ax sub \By \Ay sub){x_3}
  \pnode(! \Cx \Bx sub \Cy \By sub){x_1}
  \pnode(! \Ax \Cx sub \Ay \Cy sub){x_2}
  \ncline[linewidth=1.2pt]{*->}{X_1}{X_2}
  \Bput{$\vc{x}_3$}
  \ncline[linewidth=1.2pt]{*->}{X_2}{X_3}
  \Bput{$\vc{x}_1$}
  \ncline[linewidth=1.2pt]{*->}{X_3}{X_1}
  \Bput{$\vc{x}_2$}
  \uput[180](X_1){$\vc{X}_1$}
  \uput[-60](X_2){$\vc{X}_2$}
  \uput[90](X_3){$\vc{X}_3$}
  \psdots[linewidth=1.2pt](H)
  \psline[linestyle=dashed](X_1)(! \HAx \HAy)
  \psline[linestyle=dashed](X_2)(! \HBx \HBy)
  \psline[linestyle=dashed](X_3)(! \HCx \HCy)
  \uput[-65](H){$H$}
  \pcline[linestyle=dashed]{-}(X_1)([nodesep=2]X_1)
  \psarc{->}(X_1){1.5}{0}{(x_3)}
  \uput{1.7}[! \By \Ay sub \Bx \Ax sub atan 2 div](X_1){$\varphi_3$}
  \pcline[linestyle=dashed]{-}(X_2)([nodesep=2]X_2)
  \psarc{->}(X_2){1.5}{0}{(x_1)}
  \uput{1.7}[! \Cy \By sub \Cx \Bx sub atan 2 div](X_2){$\varphi_1$}
  \pcline[linestyle=dashed]{-}(X_3)([nodesep=2]X_3)
  \psarc{->}(X_3){1.5}{0}{(x_2)}
  \uput{1.7}[! \Ay \Cy sub \Ax \Cx sub atan 2 div](X_3){$\varphi_2$}
  \psarc(X_1){1}{(x_3)}{! \Cy \Ay sub \Cx \Ax sub atan}
  \uput{1.2}[! \By \Ay sub \Bx \Ax sub atan 
  \Cy \Ay sub \Cx \Ax sub atan add 2 div](X_1){$\phi_1$}
  \psarc(X_2){1}{(x_1)}{! \Ay \By sub \Ax \Bx sub atan}
  \uput{1.2}[! \Cy \By sub \Cx \Bx sub atan 
  \Ay \By sub \Ax \Bx sub atan add 2 div](X_2){$\phi_2$}
  \psarc(X_3){1}{(x_2)}{! \By \Cy sub \Bx \Cx sub atan}
  \uput{1.2}[! \Ay \Cy sub \Ax \Cx sub atan 
  \By \Cy sub \Bx \Cx sub atan add 2 div](X_3){$\phi_3$}
\end{pspicture}%
%
}}
  \hfill
  \parbox[b]{0.39\hsize}{%
    \caption{Definitions of the geometry of a triangle. The $\vc X_l$
      are the vertices of the triangle, the $\vc x_l$ the
      corresponding sides, $\vp_l$
      are the orientations of the sides relative to the
      positive $x_1$-direction, the $\phi_l$ are the interior angles of
      the triangle.  The figure also shows the orthocenter $H$, i.e.\
      the intersection of the heights of the triangle.}%
    \label{fig:1}}
\end{figure}
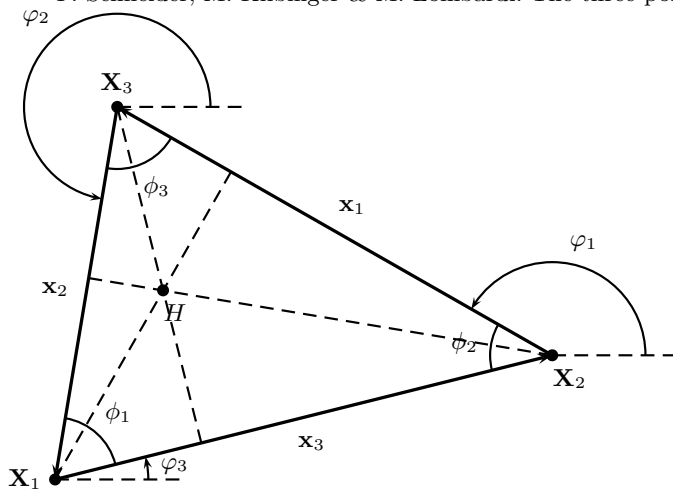

We specify our geometry in the same way as in Paper~I (see Fig.~1):
Let $\vc X_i$ be three points at which the shear is considered. The 
connecting vectors are $\vc x_1=\vc X_3-\vc X_2$, $\vc x_2=\vc X_1-\vc X_3$,
and $\vc x_3=\vc X_2-\vc X_1$. We denote the polar angle of the vector
$\vc x_i$
by $\vp_i$, and the interior angle of the triangle at the point $\vc X_i$ by
$\phi_i$. Furthermore, we assume that the triangle is oriented such that $\vc
x_1\times \vc x_2=\vc x_2\times \vc x_3=\vc x_3\times \vc x_1>0$,
where for two two-dimensional vectors $\vc a$ and $\vc b$ we defined
$\vc a\times\vc b=a_1 b_2-a_2b_1$. Hence, the points $\vc X_i$ are
ordered counter-clockwise around the triangle.

In the first part of this paper, we shall consider the projection of
the shear relative to the orthocenter of the triangle. As the vector
connecting the point $\vc X_i$ and the orthocenter is perpendicular to
the vector $\vc x_i$ (see Fig.~1), one has $\zeta_i=\vp_i+\pi/2$ in
this case, so that
\be
\gamma^{(\rm o)}(\vc X_i)\equiv 
\gamma^{(\rm o)}_{\rm t}(\vc X_i)
+{\rm i}\gamma^{(\rm o)}_\times(\vc X_i)
=\eck{\gamma_1(\vc X_i)+{\rm i}\gamma_2(\vc X_i)}\,{\rm e}^{-2{\rm
i}\vp_i}
\equiv
\gamma(\vc X_i)\,{\rm e}^{-2{\rm
i}\vp_i} \;.
\elabel{Gortho}
\ee
The shear 3PCF depends linearly on the 3PCF of the convergence, or
equivalently on its Fourier transform, the bispectrum.
The bispectrum of the surface mass density is defined as (see, e.g.,
van Waerbeke et al.\ 1999)
\be
\ave{\hat\kappa(\vc\ell_1)\hat\kappa(\vc\ell_2)\hat\kappa(\vc\ell_3)}
=(2\pi)^2\eck{B(\vc\ell_1,\vc\ell_2)
+B(\vc\ell_2,\vc\ell_3)+B(\vc\ell_3,\vc\ell_1)}
\,\delta(\vc\ell_1+\vc\ell_2+\vc\ell_3)
\;,
\elabel{bispec}
\ee
hence, it is non-zero only for closed triangles in $\ell$-space. This 
follows from the assumed statistical homogeneity of the random field
$\kappa$. Furthermore, if $\kappa$ is an isotropic random field, the function
$B(\vc\ell,\vc\ell')$ depends only on $|\vc\ell|$, $|\vc\ell'|$, and the angle
$\vp$ enclosed by $ \vc\ell$ and $\vc\ell'$. We shall therefore write
$B(\vc\ell,\vc\ell') =b(|\vc\ell|,|\vc\ell'|,\vp)$. If, in addition, the
statistical properties of the field $\kappa$ are invariant under parity
transformation (as we shall assume throughout), then $b$ is an even function
of $\vp$, or, equivalently, $b$ is invariant against exchanging $\ell$ and
$\ell'$,
\be
b(\ell,\ell',-\vp)=b(\ell,\ell',\vp)=b(\ell',\ell,\vp) \;.
\elabel{qparity}
\ee

\section{\llabel{STPCF}The shear three-point correlation function in terms of
the bispectrum}
In this section, we will derive the shear 3PCF in terms of the bispectrum $B$
of the convergence. As it turns out, the calculations are fairly cumbersome,
owing to the rich mathematical structure of the 3PCF with its three
arguments, compared to the 2PCF which has only one argument (and for which the
direction along which the shear components are measured is uniquely given by
the connecting vector between any pair of points). 

\subsection{\llabel{STPCF0}The case of $\Gamma^{(0)}$}
The natural component $\Gamma^{(0)}$ of the shear 3PCF measured
relative to the orthocenter reads
\bea
\Gamma^{(0)}(x_1,x_2,x_3)&:=&
\ave{\gamma^{(\rm o)}(\vc X_1)\gamma^{(\rm o)}(\vc X_2)
\gamma^{(\rm o)}(\vc X_3)}
=\int{\d^2\ell_1\over (2\pi)^2}\int{\d^2\ell_2\over (2\pi)^2}
\int{\d^2\ell_3\over (2\pi)^2}
\exp\eck{-{\rm i}\rund{\vc\ell_1\cdot\vc X_1
+\vc\ell_2\cdot\vc X_2+\vc\ell_3\cdot\vc X_3}} \nonumber \\
&\times& \exp\eck{2{\rm i}\rund{\sum\beta_i-\sum\vp_i}}
\ave{\hat\kappa(\vc\ell_1)\hat\kappa(\vc\ell_2)\hat\kappa(\vc\ell_3)} \;,
\elabel{Gam0-1}
\eea
where we made use of the relation (\ref{eq:FTrel}) between the Fourier
transforms of the shear and the convergence, and the definition
(\ref{eq:Gortho}) of the shear components relative to the orthocenter.
Inserting the bispectrum (\ref{eq:bispec}) into Eq.\
(\ref{eq:Gam0-1}), performing 
for each of the resulting three terms the integration over the
$\vc\ell$-vector which is not in the argument of the $B$ function, by making
use of the delta-`function' in (\ref{eq:bispec}), and using the relations
between the corner points $\vc X_i$ and the side vectors $\vc x_i$, one
obtains after renaming the dummy integration variables
\bea
\Gamma^{(0)}(x_1,x_2,x_3)&=&
\int_0^\infty{\d \ell_1\;\ell_1\over (2\pi)^2}
\int_0^\infty{\d \ell_2\;\ell_2\over (2\pi)^2}
\int_0^{2\pi}\d\beta_1
\int_0^{2\pi}\d\beta_2\;
b(\ell_1,\ell_2,\beta_1-\beta_2) \;
\exp\eck{2{\rm i}\rund{\sum\beta_i-\sum\vp_i}} \nonumber \\
&\times&
\eck{{\rm e}^{{\rm i}(\vc \ell_2\cdot\vc x_1-\vc \ell_1\cdot\vc x_2)} 
+{\rm e}^{{\rm i}(\vc \ell_2\cdot\vc x_2-\vc \ell_1\cdot\vc x_3)}
+{\rm e}^{{\rm i}(\vc \ell_2\cdot\vc x_3-\vc \ell_1\cdot\vc x_1)}} \;.
\elabel{Gam0-2}
\eea
The angle $\beta_3$ occurring in (\ref{eq:Gam0-2}) 
is the polar angle of the vector $\vc\ell_3=-\vc\ell_1-\vc\ell_2$, so that
\[
\cos2\beta_3={\ell_1^2\cos2\beta_1+\ell_2^2\cos 2\beta_2+2\ell_1\ell_2
\cos(\beta_1+\beta_2)\over \abs{\vc \ell_1+\vc \ell_2}^2}\; ;\quad
\sin2\beta_3={\ell_1^2\sin2\beta_1+\ell_2^2\sin 2\beta_2+2\ell_1\ell_2
\sin(\beta_1+\beta_2)\over \abs{\vc \ell_1+\vc \ell_2}^2}\; .
\]
We next
rename the angles in the following way:
\be
\beta_1=\theta+\vp/2\;;\quad
\beta_2=\theta-\vp/2\;;\quad
\beta_3=\theta+\bar\beta\;,
\elabel{angletrans}
\ee
so that $\vp$ is the angle between $\vc\ell_1$ and $\vc\ell_2$, as previously
defined, and $\bar\beta$ is the angle between the direction of $\vc\ell_3$ and
the mean of the directions of $\vc\ell_1$ and $\ell_2$.  Since $B$ is
independent of $\theta$, a further integration can be carried out in
(\ref{eq:Gam0-2}), using
$\int\d\beta_1\int\d\beta_2=\int\d\theta\int\d\vp$. Owing to the symmetric
form of the three terms occurring, only one of the three terms has to be
calculated explicitly; we shall consider the first term in the following. From
the foregoing equations one finds that
\be
\cos 2\bar\beta={(\ell_1^2+\ell_2^2)\cos\vp+2\ell_1\ell_2\over
\abs{\vc \ell_1+\vc \ell_2}^2} \; ;\quad
\sin 2\bar\beta={(\ell_1^2-\ell_2^2)\sin\vp\over
\abs{\vc \ell_1+\vc \ell_2}^2} \; ;\quad {\rm with}\quad
\abs{\vc\ell_1+\vc\ell_2}^2=\ell_1^2+\ell_2^2+2\ell_1\ell_2\cos\vp\;.
\elabel{bbar}
\ee
Next, we consider the argument of the exponential,
\bea
\vc\ell_2\cdot\vc x_1-\vc\ell_1\cdot \vc x_2
&=&\ell_2 x_1 \cos\rund{\theta-\vp/2-\vp_1} -\ell_1 x_2
\cos\rund{\theta+\vp/2-\vp_2} \nonumber  \\
&=&-\ell_2 x_1 \sin\eck{\theta'-\rund{\vp+\phi_3}/2}
-\ell_1 x_2 \sin\eck{\theta'+\rund{\vp+\phi_3}/2} \;,
\elabel{EllX1}
\eea
where we have defined $\theta=\theta'+(\vp_1+\vp_2)/2$ and used the 
fact that $\vp_2-\vp_1=\pi-\phi_3$. Therefore, we can write
\be
\vc\ell_2\cdot\vc x_1-\vc\ell_1\cdot \vc x_2 = -A_3
\sin(\theta'+\alpha_3) \;,
\elabel{EllX2}
\ee
from which one finds, after expanding the trigonometric functions in
(\ref{eq:EllX1}) and (\ref{eq:EllX2}),
\bea
A_3\cos\alpha_3&=&(\ell_1 x_2+\ell_2 x_1)\cos\rund{\vp+\phi_3\over 2}\;
; \quad
A_3\sin\alpha_3=(\ell_1 x_2-\ell_2 x_1)\sin\rund{\vp+\phi_3\over 2}\;
; \nonumber \\
A_3&=&\sqrt{(\ell_1 x_2)^2 + (\ell_2 x_1)^2+2\ell_1 \ell_2 x_1 x_2
\cos(\vp+\phi_3)} \;.
\elabel{Aalpha}
\eea
Finally, we consider the sums over the angles that occur in
(\ref{eq:Gam0-2}), 
\be
\sum\beta_i-\sum\vp_i=3\theta+\bar\beta-\sum\vp_i
=3\theta'+\bar\beta+(\vp_1+\vp_2)/2-\vp_3
=3\theta'+\bar\beta+(\phi_1-\phi_2)/2 \;.
\ee
We can now perform the $\theta$-integration of the first term in
(\ref{eq:Gam0-2}) as follows:
\bea
\int_0^{2\pi}\!\!\!\!\!\!\!\!
&&\d\theta\;\exp\eck{2{\rm i}\rund{\sum\beta_i-\sum\vp_i}}
{\rm e}^{{\rm i}(\vc\ell_2\cdot \vc x_1-\vc\ell_1\cdot \vc x_2)}
={\rm e}^{2{\rm i}\bar\beta}\,{\rm e}^{{\rm i}(\phi_1-\phi_2)}
\int_0^{2\pi}\d\theta'\;{\rm e}^{6{\rm i}\theta'}\,
{\rm e}^{-{\rm i}A_3\sin(\theta'+\alpha_3)} \nonumber \\
&=&{\rm e}^{2{\rm i}\bar\beta}\,{\rm e}^{{\rm i}(\phi_1-\phi_2)}
\int_0^{2\pi}\d\vt\;{\rm e}^{6{\rm i}(\vt-\alpha_3)}\,
{\rm e}^{-{\rm i}A_3\sin\vt}
=2\pi \;{\rm e}^{2{\rm i}\bar\beta}\,{\rm e}^{{\rm i}(\phi_1-\phi_2)}\,
{\rm e}^{-6{\rm i}\alpha_3}\; {\rm J}_6(A_3) \;,
\eea
where ${\rm J}_n$ is the Bessel function of the first kind. Therefore,
(\ref{eq:Gam0-2}) becomes
\bea
\Gamma^{(0)}(x_1,x_2,x_3)&=&
(2\pi) \int_0^\infty{\d \ell_1\;\ell_1\over (2\pi)^2}
\int_0^\infty{\d \ell_2\;\ell_2\over (2\pi)^2}
\int_0^{2\pi}\d\vp\;
b(\ell_1,\ell_2,\vp) \;{\rm e}^{2{\rm i}\bar\beta}\nonumber \\
&\times& \eck{{\rm e}^{{\rm i}(\phi_1-\phi_2-6\alpha_3)}{\rm J}_6(A_3)
+{\rm e}^{{\rm i}(\phi_2-\phi_3-6\alpha_1)}{\rm J}_6(A_1)
+{\rm e}^{{\rm i}(\phi_3-\phi_1-6\alpha_2)}{\rm J}_6(A_2)} \;,
\elabel{Gam0}
\eea
where the $A_i$ and $\alpha_i$ are obtained from (\ref{eq:Aalpha}) by
cyclic permutations of the $x_1,x_2,x_3$.

\subsection{\llabel{STPCF1}The case of $\Gamma^{(1)}$}
Next, we calculate the natural component 
\bea
\Gamma^{(1)}(x_1,x_2,x_3)&:=&
\ave{\gamma^{(\rm o)*}(\vc X_1)\gamma^{(\rm o)}(\vc X_2)
\gamma^{(\rm o)}(\vc X_3)}
=\int{\d^2\ell_1\over (2\pi)^2}\int{\d^2\ell_2\over (2\pi)^2}
\int{\d^2\ell_3\over (2\pi)^2}
\exp\eck{{\rm i}\rund{\vc\ell_1\cdot\vc X_1
-\vc\ell_2\cdot\vc X_2-\vc\ell_3\cdot\vc X_3}} \nonumber \\
&\times& {\rm e}^{2{\rm i}\rund{\beta_2+\beta_3-\beta_1+\vp_1-\vp_2-\vp_3}}
\ave{\hat\kappa(-\vc\ell_1)\hat\kappa(\vc\ell_2)\hat\kappa(\vc\ell_3)} \;,
\elabel{Gam1-1}
\eea
where we made use of the fact that
$\hat\kappa^*(\vc\ell)=\hat\kappa(-\vc\ell)$, since $\kappa(\vc x)$ is
a real field. Next, we change the integration variable $\vc\ell_1 \to
-\vc \ell_1$; as a consequence, $\beta_1 \to \beta_1+\pi$, but this
does not change the exponential in (\ref{eq:Gam1-1}). Inserting the
bispectrum in the form (\ref{eq:bispec}), and appropriately renaming
the dummy integration variables, one finds
\bea
\Gamma^{(1)}(x_1,x_2,x_3)&=&{\rm e}^{2{\rm i}\rund{\vp_1-\vp_2-\vp_3}}
\int_0^\infty{\d \ell_1\;\ell_1\over (2\pi)^2}
\int_0^\infty{\d \ell_2\;\ell_2\over (2\pi)^2}
\int_0^{2\pi}\d\beta_1
\int_0^{2\pi}\d\beta_2\;
b(\ell_1,\ell_2,\beta_1-\beta_2) \;
 \nonumber \\
&\times&
\eck{{\rm e}^{{\rm i}(\vc \ell_2\cdot\vc x_1-\vc \ell_1\cdot\vc x_2)}
 {\rm e}^{2{\rm i}(\beta_2+\beta_3-\beta_1)}
+{\rm e}^{{\rm i}(\vc \ell_2\cdot\vc x_2-\vc \ell_1\cdot\vc x_3)}
 {\rm e}^{2{\rm i}(\beta_1+\beta_2-\beta_3)}
+{\rm e}^{{\rm i}(\vc \ell_2\cdot\vc x_3-\vc \ell_1\cdot\vc x_1)} 
 {\rm e}^{2{\rm i}(\beta_3+\beta_1-\beta_2)}}\;.
\elabel{Gam1-2}
\eea
The further calculation proceeds in the same way as in the case of
$\Gamma^{(0)}$. Specifically, we employ the change of angular
integration variables given in (\ref{eq:angletrans}), evaluate the
three exponentials containing products of the form $\vc
x_i\cdot\vc\ell_j$ using (\ref{eq:EllX1}), (\ref{eq:EllX2}), and their
analogous expressions obtained by cyclic permutations of the $x_i$,
and calculating the angular sums in the exponentials. The final result
reads
\bea
\Gamma^{(1)}(x_1,x_2,x_3)&=& 2\pi 
\int_0^\infty{\d \ell_1\;\ell_1\over (2\pi)^2}
\int_0^\infty{\d \ell_2\;\ell_2\over (2\pi)^2}
\int_0^{2\pi} \d\vp \;b(\ell_1,\ell_2,\vp)
\;\biggl[ {\rm e}^{{\rm i}(\phi_1-\phi_2+2\phi_3)}
{\rm e}^{2{\rm i}(\bar\beta-\vp-\alpha_3)}\;{\rm J}_2(A_3) \nonumber \\
&+& 
{\rm e}^{{\rm i}(\phi_3-\phi_2)}
{\rm e}^{-2{\rm i}(\bar\beta+\alpha_1)}\;{\rm J}_2(A_1)
+{\rm e}^{{\rm i}(\phi_3-\phi_1-2\phi_2)}
{\rm e}^{2{\rm i}(\bar\beta+\vp-\alpha_2)}\;{\rm J}_2(A_2) \biggr] \;.
\elabel{Gam1}
\eea
The expressions for the other two natural components $\Gamma^{(2)}$
and $\Gamma^{(3)}$ of the shear 3PCF, which are defined in analogy to
(\ref{eq:Gam1-1}) by placing the complex conjugation of the shear at
point $\vc X_2$ and $\vc X_3$, respectively, are obtained from
(\ref{eq:Gam1}) by applying the transformation laws given in Paper\ts
I, i.e., even permutations of the arguments.

The resulting expressions (\ref{eq:Gam0}) and (\ref{eq:Gam1}) are not only
relatively complicated, but their numerical evaluation also is quite
cumbersome. Recalling that the relation between the 2PCF of the shear and the
power spectrum $P_\kappa(\ell)$ of the convergence involves a convolution
integral over a Bessel function, one should perhaps not be too surprised that
in the case of third-order statistics there are three such oscillating factors
in the transformation between the shear 3PCF and the bispectrum. In a
future work, we will investigate numerical procedures with which the
integration can be carried out accurately; first attempts, using
Gaussian quadrature for the two $\ell$-integrations and an equidistant
grid for the $\vp$-integration already yielded satisfactory
results. Hence, despite the apparent complexity, the foregoing
equations can be applied in practice.

\section{\llabel{Transfo}Transformation laws}
In Paper\ts I the behavior of the natural components under a change of
the order 
of the arguments was discussed, using simple geometrical arguments.  We shall
now consider this behavior explicitly, using the expressions (\ref{eq:Gam0})
and (\ref{eq:Gam1}). First, consider $\Gamma^{(0)}$.

Taking an even permutation of the arguments of $\Gamma^{(0)}$ just
changes the order of the terms in the integral of (\ref{eq:Gam0}) and
therefore leaves $\Gamma^{(0)}$ unchanged. Taking an odd permutation
of the arguments means that two of the arguments are interchanged,
e.g., $x_1$ and $x_2$.  Interchanging $x_1$ and $x_2$ corresponds to an
interchange of $\phi_1$ and $\phi_2$. Using the property
(\ref{eq:qparity}), one can also interchange $\ell_1$ and
$\ell_2$. From (\ref{eq:bbar}) one sees that these changes imply that
$\bar\beta \to -\bar\beta$. Furthermore, from (\ref{eq:Aalpha}), one
sees that these changes lead to $A_3 \to A_3$, and $\alpha_3 \to
-\alpha_3$. Together this implies that these transformations lead to a
complex conjugation of the first term in (\ref{eq:Gam0}). From the
expressions for $A_l$ and $\alpha_l$ obtained from (\ref{eq:Aalpha})
by cyclic permutations of the $x_k$, one finds that the above
interchanges of $x_1$ and $x_2$ leads to $A_2 \to A_1$, $A_1 \to A_2$,
$\alpha_2 \to -\alpha_1$, $\alpha_1 \to -\alpha_2$. This then implies
that the second term in (\ref{eq:Gam0}) becomes the complex conjugate
of the third term, and vice versa. Taken together, we see that an odd
permutation of the arguments changes $\Gamma^{(0)} \to
\Gamma^{(0)*}$, as already argued from parity considerations in
Paper\ts I.

Note that an odd permutation of the arguments in geometric terms means that
the orientation of the triangle is reversed. We shall now show that this is
equivalent to replacing all $\phi_l$ by $-\phi_l$. The motivation for this
observation comes from the fact that for a triangle with odd orientation, the
relations [see Eq.~(1) of Paper~I] between the orientations $\vp_i$ of the
$\vc x_i$ and the interior angles $\phi_i$ formally yield
$\phi_i\in[-\pi,0]$ (modulo $2\pi$), whereas for a triangle with even
orientation, $\phi_i\in[0,\pi]$ (modulo $2\pi$). If we apply this
transformation, $\phi_i\to -\phi_i$, we can change the integration variable
$\vp \to -\vp$ in (\ref{eq:Gam0}), noting from (\ref{eq:qparity}) that
$b(\ell_1,\ell_2,\vp)$ is unaffected by this change. These two changes
together then imply that $\bar\beta \to -\bar\beta$, $A_i\to A_i$, and
$\alpha_i \to -\alpha_i$ [see (\ref{eq:bbar}) and (\ref{eq:Aalpha}),
respectively]. Hence, all three terms of (\ref{eq:Gam0}) are transformed to
their complex conjugates, as was claimed above. Note that this transformation
behavior directly implies that $\Gamma^{(0)}$ is real if two of its arguments
are equal.

Next one can consider the transformation behavior of
$\Gamma^{(1)}$. Cyclic permutations of the arguments transform
$\Gamma^{(1)}$ into $\Gamma^{(2)}$ and $\Gamma^{(3)}$, yielding the
transformation behavior derived in Paper\ts I. Interchanging $x_2$ and
$x_3$ (and thus $\phi_2$ and $\phi_3$)
should yield the complex conjugate of $\Gamma^{(1)}$. Again, we
interchange $\ell_1$ and $\ell_2$, which then yields $\bar\beta \to
-\bar\beta$, $A_1\to A_1$, $\alpha_1 \to -\alpha_1$, and so the second
term in (\ref{eq:Gam1}) is complex conjugated. Furthermore, these
transformations yield $A_2 \to A_3$, $\alpha_2 \to -\alpha_3$,  
$A_3 \to A_2$, $\alpha_3 \to -\alpha_2$, which shows that the first
term in (\ref{eq:Gam1}) becomes the complex conjugate of the third,
and vice versa, so that $\Gamma^{(1)} \to \Gamma^{(1)*}$, as was to be
shown. This transformation implies that $\Gamma^{(1)}$ is real if the
last two arguments are equal.

Equivalently, we can also let $\phi_l \to -\phi_l$, and change the
integration variable $\vp \to -\vp$. This implies $A_l \to A_l$,
$\alpha_l \to -\alpha_l$, $\bar\beta \to -\bar\beta$, and each term in
(\ref{eq:Gam1}) is transformed into its complex conjugate.

\section{\llabel{Bispect}Bispectrum in terms of the 3PCF}
Recall the situation for the two-point correlation of the shear: there, the
relation between the correlation functions and the power spectrum of the
projected density fluctuations can be inverted (e.g., Schneider et al.\ 2002;
hereafter SvWM), and thus the power spectrum can be expressed in terms of the
two-point correlation function. It will be shown here that in analogy, the
bispectrum $B(\vc \ell_1,\vc\ell_2)$ can be expressed in terms of the 3PCF.

\subsection{\llabel{BispG0}Bispectrum in terms of $\Gamma^{(0)}$}
From (\ref{eq:FTrel}) one finds that
\bea
\ave{\hat\kappa(\vc\ell_1)\hat\kappa(\vc\ell_2)\hat\kappa(\vc\ell_3)}
&=&
{\rm e}^{-2{\rm i}\sum\beta_i}\ave{\hat\gamma(\vc\ell_1)\hat\gamma(\vc\ell_2)
\hat\gamma(\vc\ell_3)} \nonumber \\
=
{\rm e}^{-2{\rm i}\sum\beta_i}
\!\!\!\!\!&&\!\!\!\!\!
\int\d^2 X_1\int\d^2 X_2 \int \d^2 X_3 \;
{\rm e}^{2{\rm i}\sum\vp_i}\,{\rm e}^{{\rm i}(\vc \ell_1\cdot\vc X_1
+\vc \ell_2\cdot\vc X_2 + \vc \ell_3\cdot\vc X_3)}
\ave{\gamma^{(\rm o)}(\vc X_1)\gamma^{(\rm o)}(\vc X_2)\gamma^{(\rm
o)}(\vc X_3)} \;,
\eea
where in the second step we used (\ref{eq:Gortho}). We now split the
right-hand side of the foregoing equation into three identical terms,
each of which is thus one third of the above expression, and
substitute $\vc X_1=\vc X_3+\vc x_2$, $\vc X_2=\vc X_3-\vc x_1$ in the
first of these, and similar substitutions, obtained by cyclic
permutations of the $\vc X_l$ and $\vc x_l$ in the other two
terms. This then yields
\be
\ave{\hat\kappa(\vc\ell_1)\hat\kappa(\vc\ell_2)\hat\kappa(\vc\ell_3)}
={1\over 3} {\rm e}^{-2{\rm i}\sum\beta_i}
\!\int\!\d^2 x_1\!\int\!\d^2 x_2 \!\int\! \d^2 X_3 \;
{\rm e}^{2{\rm i}\sum\vp_i}
{\rm e}^{{\rm i}(\vc \ell_1\cdot \vc x_2-\vc \ell_2\cdot \vc x_1)}
{\rm e}^{{\rm i}\vc X_3\cdot(\vc \ell_1+\vc\ell_2+\vc\ell_3)}
\Gamma^{(0)}(\vc x_1,\vc x_2) + \hbox{2 terms} \;,
\elabel{Q-CF-0}
\ee
where we have used the notation $\Gamma^{(0)}(\vc x_1,\vc x_2)$ for
the 3PCF $\Gamma^{(0)}(x_1,x_2,x_3)$, with $x_3=|\vc x_1+\vc x_2|$, if
$\vc x_1\times \vc x_2\geq 0$, and $\Gamma^{(0)*}(x_1,x_2,x_3)$ if $\vc
x_1\times \vc x_2<0$. Hence, if expressed in terms of the arguments
$(\vc x_1,\vc x_2)$, the information about the orientation of the
three points $\vc X_l$ is included. Another notation to be used
later on is $\Gamma^{(0)}(x_1,x_2,\phi_3)$, which also includes the
orientation of the three points. The $\vc X_3$-integration in the
previous equation yields a delta-function; comparison with
(\ref{eq:bispec}) then results in 
\be
B(\vc \ell_1,\vc\ell_2)={1\over 3} {\rm e}^{-2{\rm i}\sum\beta_i}
\int\d^2 x_1\int\d^2 x_2\;
{\rm e}^{2{\rm i}\sum\vp_i}
{\rm e}^{{\rm i}(\vc \ell_1\cdot \vc x_2-\vc \ell_2\cdot \vc x_1)}\,
\Gamma^{(0)}(\vc x_1,\vc x_2) \;.
\elabel{Q-CF-1}
\ee
It is easy to show that (\ref{eq:Q-CF-1}) is compatible with
(\ref{eq:Gam0-2}), since
\bea
\Gamma^{(0)}(\vc x_1,\vc x_2)&=&\!\int{\d^2\ell_1\over
(2\pi)^2}\!\int\! {\d^2\ell_2\over (2\pi)^2} \,
B(\vc \ell_1,\vc\ell_2)\,
\exp\!\eck{2{\rm i}\rund{\sum\beta_i-\sum\vp_i}}
\eck{{\rm e}^{{\rm i}(\vc \ell_2\cdot\vc x_1-\vc \ell_1\cdot\vc x_2)} 
+{\rm e}^{{\rm i}(\vc \ell_2\cdot\vc x_2-\vc \ell_1\cdot\vc x_3)}
+{\rm e}^{{\rm i}(\vc \ell_2\cdot\vc x_3-\vc \ell_1\cdot\vc x_1)}}
\nonumber \\
&=&{1\over 3} \exp\rund{{-2{\rm i}\sum\vp_i}}
\int{\d^2\ell_1\over
(2\pi)^2}\int{\d^2\ell_2\over (2\pi)^2}
\int\d^2 y_1\int \d^2 y_2 
\exp\rund{2{\rm i}\sum\vp'_i}
{\rm e}^{{\rm i}(\vc \ell_1\cdot\vc y_2-\vc \ell_2\cdot\vc y_1)} 
\\ &\times &
\eck{{\rm e}^{{\rm i}(\vc \ell_2\cdot\vc x_1-\vc \ell_1\cdot\vc x_2)} 
+{\rm e}^{{\rm i}(\vc \ell_2\cdot\vc x_2-\vc \ell_1\cdot\vc x_3)}
+{\rm e}^{{\rm i}(\vc \ell_2\cdot\vc x_3-\vc \ell_1\cdot\vc x_1)}}
\Gamma^{(0)}(\vc y_1,\vc y_2)\;, \nonumber 
\eea
where the $\vp_i'$ are the polar angles of the $\vc y_i$, and $\vc
y_3=-\vc y_1-\vc y_2$. The $\vc\ell_i$-integrations can be carried
out, yielding delta-`functions': for the first term, e.g., they yield 
$(2\pi)^2\delta(\vc y_2-\vc x_2) (2\pi)^2 \delta(\vc x_1-\vc y_1)$, so
that also $\sum\vp_i=\sum\vp'_i$. Together, this yields
\be
\Gamma^{(0)}(\vc x_1,\vc x_2)={1\over 3}
\eck{\Gamma^{(0)}(\vc x_1,\vc x_2) + \Gamma^{(0)}(\vc x_2,\vc x_3) +
\Gamma^{(0)}(\vc x_3,\vc x_1)} \;,
\ee
but since $\Gamma^{(0)}(\vc x_1,\vc x_2)=\Gamma^{(0)}(\vc x_2,\vc x_3)
=\Gamma^{(0)}(\vc x_3,\vc x_1)$, the compatibility of
(\ref{eq:Q-CF-1}) and (\ref{eq:Gam0-2}) has been shown.

Next, we want to calculate $b(\ell_1,\ell_2,\vp)$ from
(\ref{eq:Q-CF-1}), i.e. taking a further integration. For that
purpose, we write $\vp_1=\mu+\psi/2$, $\vp_2=\mu-\psi/2$, where
$\vp_1-\vp_2=\psi=\phi_3-\pi$. Then,
\bea
\vc \ell_1\cdot\vc x_2 -\vc\ell_2\cdot\vc x_1
&=&\ell_1 x_2 \cos(\theta+\vp/2-\mu+\psi/2)
-\ell_2 x_1 \cos(\theta-\vp/2-\mu-\psi/2) \nonumber \\
&=& -\ell_1 x_2 \sin\eck{\mu'-(\vp+\phi_3)/2} 
-\ell_2 x_1\sin\eck{\mu'+(\vp+\phi_3)/2}
=-A_3 \sin(\mu'-\alpha_3)\;,
\eea
where we defined $\mu'=\mu-\theta$ in the second step, and $A_3$ and
$\alpha_3$ are given in (\ref{eq:Aalpha}). Writing the polar angle of
$\vc x_3$ as $\vp_3=\mu+\bar\vp$, one finds that 
$\sum\vp_i-\sum\beta_i=3\mu'+\bar\vp-\bar\beta$, where
\be
\cos 2\bar\vp={-(x_1^2+x_2^2)\cos\phi_3+2x_1x_2\over
\abs{\vc x_1+\vc x_2}^2} \; ;\quad
\sin 2\bar\vp=-{(x_1^2-x_2^2)\sin\phi_3 \over \abs{\vc x_1+\vc x_2}^2}
\;,
\ee
and $\bar\beta$ is given in (\ref{eq:bbar}). Taken together,
(\ref{eq:Q-CF-1}) becomes 
\bea
b(\ell_1,\ell_2,\vp)&=&{1\over 3} {\rm e}^{-2{\rm i}\bar\beta}
\int_0^\infty \d x_1\,x_1
\int_0^\infty \d x_2\,x_2
\int_0^{2\pi}\d\phi_3\;
\Gamma^{(0)}(x_1,x_2,\phi_3) \,{\rm e}^{2{\rm i}\bar\vp}
\int_0^{2\pi} \d\mu' {\rm e}^{6{\rm i}\mu'}\,
{\rm e}^{-{\rm i}A_3\sin(\mu'-\alpha_3)} \nonumber \\
&=&
{2\pi\over 3} {\rm e}^{-2{\rm i}\bar\beta}
\int_0^\infty \d x_1\,x_1
\int_0^\infty \d x_2\,x_2
\int_0^{2\pi}\d\phi_3\;
\Gamma^{(0)}(x_1,x_2,\phi_3) \,{\rm e}^{2{\rm i}\bar\vp}
\, {\rm e}^{6{\rm i}\alpha_3}\,{\rm J}_6(A_3)  \;.
\elabel{qsmall}
\eea
It is easy to see that $b(\ell_1,\ell_2,\vp)$ as given by
(\ref{eq:qsmall}) is real: Since $b^*(\ell_1,\ell_2,\vp)=
b^*(\ell_1,\ell_2,-\vp)$ -- see (\ref{eq:qparity}) -- and
$\Gamma^{(0)*}(x_1,x_2,\phi_3)=\Gamma^{(0)}(x_1,x_2,-\phi_3)$, taking
the complex conjugate of (\ref{eq:qsmall}) and simultaneously
replacing $\vp \to -\vp$, and the integration variable $\phi_3\to
-\phi_3$ yields the same expression as (\ref{eq:qsmall}),
$b^*(\ell_1,\ell_2,\vp)=b(\ell_1,\ell_2,\vp)$.

\def\ii{{\rm i}}
\def\dd{{\rm d}}
\subsection{\llabel{BispG1}Bispectrum in terms of the other $\Gamma^{(i)}$}
Next we calculate the bispectrum as a function of the other three natural
components of the shear 3PCF, starting from
\begin{eqnarray}
\ave{\hat\kappa_1(\vc \ell_1) \hat\kappa_1(\vc \ell_2)
\hat\kappa_1(\vc \ell_3)} & = & \ave{\hat\kappa_1^*(-\vc \ell_1)
\hat\kappa_1(\vc \ell_2) \hat\kappa_1(\vc \ell_3)} \nonumber\\
 =  {\rm e}^{-2 \ii (-\beta_1 + \beta_2 + \beta_3)}
  \!\!\!\!\!   &&\!\!\!\!\int\! \dd^2 X_1\!\! \int\! \dd^2 X_2
  \!\!\int\! \dd^2 X_3 \,
        {\rm e}^{\ii \left( \vc \ell_1 \cdot \vc X_1 + \vc \ell_2 \cdot \vc X_2 +
        \vc \ell_3 \cdot \vc X_3 \right)} 
 {\rm e}^{2 \ii (-\vp_1 + \vp_2 + \vp_3)}
        \ave{{\gamma^{(\rm o)*}}(\vc X_1) \gamma^{(\rm o)}(\vc
        X_2) \gamma^{(\rm o)}(\vc X_3)}
\end{eqnarray}
where we used
\begin{equation}
\hat\gamma^*(-\vc \ell) = \int \dd^2 X\; {\rm e}^{{\rm i} \vc \ell \cdot \vc X}
{\gamma^{(\rm o)*}}(\vc X)\, {\rm e}^{-2\ii\vp}
\end{equation}
In complete analogy to (\ref{eq:Q-CF-1}), we split the right-hand side into
three terms and make appropriate substitutions. From a comparison with 
(\ref{eq:bispec}) we then obtain
\begin{eqnarray}
& & B(\vc \ell_1, \vc \ell_2) + B(\vc \ell_2, \vc \ell_3) + B(\vc
        \ell_3, \vc \ell_1)
 =  \frac 1 3 {\rm e}^{-2 \ii (-\beta_1 + \beta_2 + \beta_3)}
        \left[ \int \dd^2 x_1 \int \dd^2 x_2 \;
        {\rm e}^{{\rm i} (\vc \ell_1 \cdot \vc x_2 - \vc \ell_2 \cdot \vc x_1)}
        \right.\nonumber\\
& & \quad       + \int \dd^2 x_2 \int \dd^2 x_3\;
        {\rm e}^{{\rm i} (\vc \ell_2 \cdot \vc x_3 - \vc \ell_3 \cdot \vc x_1)}
        \left. + \int \dd^2 x_1 \int \dd^2 x_3 \;
        {\rm e}^{{\rm i} (\vc \ell_3 \cdot \vc x_1 - \vc \ell_1 \cdot \vc x_3)}
        \right]
        \Gamma^{(1)}(\vc x_1, \vc x_2)\, {\rm e}^{2 \ii (-\vp_1 + \vp_2 +
        \vp_3)}. 
\end{eqnarray}
Here, we wrote the 3PCF as 
$\ave{{\gamma^{(\rm o)*}}(\vc X_1) \gamma^{(\rm o)}(\vc
        X_2) \gamma^{(\rm o)}(\vc X_3)} \equiv \Gamma^{(1)}(\vc x_1, \vc x_2)$.
When renaming the integration variables, we have to apply the
transformation rules to the 3PCFs (see Paper~I). For the second term, we
perform the substitutions $\vc x_2 \to \vc x_1, \vc x_3
\to \vc x_2$, so that $\Gamma^{(1)}(\vc x_1, \vc x_2)
\to \Gamma^{(1)}(\vc x_3, \vc x_1) = \Gamma^{(3)}(\vc x_1,
\vc x_2)$. The third term is transformed similarly, and we get
\begin{eqnarray}
B(\vc \ell_1, \vc \ell_2) + B(\vc \ell_2, \vc \ell_3) \!\!&+&\!\! B(\vc
        \ell_3, \vc \ell_1)
= \frac 1 3 {\rm e}^{-2 \ii (-\beta_1 + \beta_2 + \beta_3)}
        \int \dd^2 x_1 \int \dd^2 x_2 
        \left[ {\rm e}^{2 \ii (-\vp_1 + \vp_2 + \vp_3)}
        {\rm e}^{\ii (\vc \ell_1 \cdot \vc x_2 - \vc \ell_2 \cdot \vc x_1)}
        \Gamma^{(1)}(\vc x_1, \vc x_2) \right. \nonumber\\
&+ &     {\rm e}^{2 \ii (\vp_1 + \vp_2 - \vp_3)}
        {\rm e}^{\ii (\vc \ell_2 \cdot \vc x_2 - \vc \ell_3 \cdot \vc x_1)}
        \Gamma^{(3)}(\vc x_1, \vc x_2) 
 +      \left. {\rm e}^{2 \ii (\vp_1 - \vp_2 + \vp_3)}
        {\rm e}^{\ii (\vc \ell_3 \cdot \vc x_2 - \vc \ell_1 \cdot \vc x_1)}
        \Gamma^{(2)}(\vc x_1, \vc x_2) \right].
\elabel{Q1}
\end{eqnarray}
Unfortunately, the three terms on the right-hand side are
not equal, as was the case for (\ref{eq:Q-CF-0}). Therefore, we repeat
the above procedure with $\ave{\hat\kappa_1(\vc \ell_1)
\hat\kappa_1^*(-\vc \ell_2) \hat\kappa_1(\vc \ell_3)}$ and
$\ave{\hat\kappa_1(\vc \ell_1) \hat\kappa_1(\vc \ell_2)
\hat\kappa_1^*(-\vc \ell_3)}$.
The two resulting equations are
\begin{eqnarray}
B(\vc \ell_1, \vc \ell_2) + B(\vc \ell_2, \vc \ell_3)\!\!& +&\!\! B(\vc
        \ell_3, \vc \ell_1)
 =  \frac 1 3 {\rm e}^{-2 \ii (\beta_1 - \beta_2 + \beta_3)}
        \int \dd^2 x_1 \int \dd^2 x_2\,
         \left[ {\rm e}^{2 \ii (\vp_1 - \vp_2 + \vp_3)}
        {\rm e}^{\ii (\vc \ell_1 \cdot \vc x_2 - \vc \ell_2 \cdot \vc x_1)}
        \Gamma^{(2)}(\vc x_1, \vc x_2) \right. \nonumber\\
& +&     {\rm e}^{2 \ii (-\vp_1 + \vp_2 + \vp_3)}
        {\rm e}^{\ii (\vc \ell_2 \cdot \vc x_2 - \vc \ell_3 \cdot \vc x_1)}
        \Gamma^{(1)}(\vc x_1, \vc x_2) 
+       \left. {\rm e}^{2 \ii (\vp_1 + \vp_2 - \vp_3)}
        {\rm e}^{\ii (\vc \ell_3 \cdot \vc x_2 - \vc \ell_1 \cdot \vc x_1)}
        \Gamma^{(3)}(\vc x_1, \vc x_2) \right] \;,
\elabel{Q2}
\end{eqnarray}
and
\begin{eqnarray}
B(\vc \ell_1, \vc \ell_2) + B(\vc \ell_2, \vc \ell_3) \!\!&+&\!\! B(\vc
        \ell_3, \vc \ell_1)
 =  \frac 1 3 {\rm e}^{-2 \ii (\beta_1 + \beta_2 - \beta_3)}
        \int \dd^2 x_1 \int \dd^2 x_2 
        \, \left[ {\rm e}^{2 \ii (\vp_1 + \vp_2 - \vp_3)}
        {\rm e}^{\ii (\vc \ell_1 \cdot \vc x_2 - \vc \ell_2 \cdot \vc x_1)}
        \Gamma^{(3)}(\vc x_1, \vc x_2) \right. \nonumber\\
& +&     {\rm e}^{2 \ii (\vp_1 - \vp_2 + \vp_3)}
        {\rm e}^{\ii (\vc \ell_2 \cdot \vc x_2 - \vc \ell_3 \cdot \vc x_1)}
        \Gamma^{(2)}(\vc x_1, \vc x_2) 
+       \left. {\rm e}^{2 \ii (-\vp_1 + \vp_2 + \vp_3)}
        {\rm e}^{\ii (\vc \ell_3 \cdot \vc x_2 - \vc \ell_1 \cdot \vc x_1)}
        \Gamma^{(1)}(\vc x_1, \vc x_2) \right].
\elabel{Q3}
\end{eqnarray}
Now, we can sum equations (\ref{eq:Q1}--\ref{eq:Q3}), after moving the
$\beta$-phase factors to the left-hand side. Then, we indeed get three
equal terms, therefore
\begin{eqnarray}
B(\vc \ell_1, \vc \ell_2) & = & \frac{1}{3g(\beta_1, \beta_2, \beta_3)}
        \int \dd^2 x_1 \int \dd^2 x_2 \;
        {\rm e}^{\ii (\vc \ell_1 \cdot \vc x_2 - \vc \ell_2 \cdot \vc x_1)}
        \nonumber \\
        &\times& \left[ {\rm e}^{2\ii(-\vp_1 + \vp_2 + \vp_3)}
        \Gamma^{(1)}(\vc x_1, \vc x_2) 
        \right. 
 + {\rm e}^{2\ii(\vp_1 - \vp_2 + \vp_3)} \Gamma^{(2)}(\vc x_1,
        \vc x_2)
        \left. + {\rm e}^{2\ii(\vp_1 + \vp_2 - \vp_3)} \Gamma^{(3)}(\vc x_1,
        \vc x_2) 
        \right]
\elabel{Qgamma1}
\end{eqnarray}
with
\begin{equation}
g(\beta_1, \beta_2, \beta_3) = {\rm e}^{2 \ii (-\beta_1 + \beta_2 +
\beta_3)} + {\rm e}^{2 \ii (\beta_1 - \beta_2 + \beta_3)} + {\rm e}^{2 \ii
(\beta_1 + \beta_2 - \beta_3)}.
\end{equation}

Equation (\ref{eq:Qgamma1}) can be written as a function of $\Gamma^{(1)}$
only, again using the transformation properties of the 3PCFs $\Gamma^{(2)}(\vc
x_1, \vc x_2) = \Gamma^{(1)}(\vc x_2, -\vc x_1 - \vc x_2)$ and
$\Gamma^{(3)}(\vc x_1, \vc x_2) = \Gamma^{(1)}(-\vc x_1 - \vc x_2, \vc x_1)$.


We use (\ref{eq:Gam1-2}) and its counterparts for $\Gamma^{(2)}$ and
$\Gamma^{(3)}$ and insert them into (\ref{eq:Qgamma1}). The $\vp$-phase
factors cancel, so that one obtains
\begin{eqnarray}
 B(\vc \ell_1, \vc \ell_2) &=& \frac{1}{3g(\beta_1, \beta_2, \beta_3)}
        \int \dd^2 x_1 \int \dd^2 x_2\;
        {\rm e}^{\ii (\vc \ell_1 \cdot \vc x_2 - \vc \ell_2 \cdot \vc x_1)}
         \int \frac{\dd^2 k_1}{(2\pi)^2} \int \frac{\dd^2 k_2}{(2\pi)^2}\;
        B(\vc k_1, \vc k_2) \nonumber\\
&\times&\Biggl[ {\rm e}^{\ii (\vc k_2 \cdot \vc x_1 - \vc k_1 \cdot \vc x_2)} {\rm e}^{2\ii\lambda_{-++}}
        + {\rm e}^{\ii (\vc k_2 \cdot \vc x_2 - \vc k_1 \cdot \vc x_3)} {\rm e}^{2\ii \lambda_{++-}}
        + {\rm e}^{\ii (\vc k_2 \cdot \vc x_3 - \vc k_1 \cdot \vc x_1)} {\rm e}^{2\ii \lambda_{+-+}}
        \nonumber\\
&& {}+ {\rm e}^{\ii (\vc k_2 \cdot \vc x_2 - \vc k_1 \cdot \vc x_3)} {\rm e}^{2\ii\lambda_{-++}}
        + {\rm e}^{\ii (\vc k_2 \cdot \vc x_3 - \vc k_1 \cdot \vc x_1)} {\rm e}^{2\ii \lambda_{++-}}
        + {\rm e}^{\ii (\vc k_2 \cdot \vc x_1 - \vc k_1 \cdot \vc x_2)} {\rm e}^{2\ii \lambda_{+-+}}
        \nonumber\\
&& {}+ {\rm e}^{\ii (\vc k_2 \cdot \vc x_3 - \vc k_1 \cdot \vc x_1)} {\rm e}^{2\ii\lambda_{-++}}
        + {\rm e}^{\ii (\vc k_2 \cdot \vc x_1 - \vc k_1 \cdot \vc x_2)} {\rm e}^{2\ii \lambda_{++-}}
        + {\rm e}^{\ii (\vc k_2 \cdot \vc x_2 - \vc k_1 \cdot \vc x_3)} {\rm
        e}^{2\ii \lambda_{+-+}} \Biggr] \;,
\elabel{Qgamma1ver}
\end{eqnarray}
with $\lambda_{\pm , \pm , \pm } \equiv \pm \lambda_1 \pm \lambda_2
\pm \lambda_3$, where $\lambda_i$ is the polar angle of the vector $\vc
k_i$. The $x_1$- and $x_2$-integrals can be performed and yield
$\delta$-functions. This makes the $k_1$- and $k_2$-integrals trivial, 
yielding $\vc k_i = \vc \ell_i$ and $\lambda_i =
\beta_i$. All the phase exponentials add up to give $3 g$, canceling the
pre-factor in (\ref{eq:Qgamma1ver}), leaving only $B(\vc \ell_1, \vc \ell_2)$ on
the right hand side and thus verifying (\ref{eq:Q3}).


As was the case for (\ref{eq:qsmall}), one additional angular integral can be
carried out. With
\begin{equation}
g(\beta_1, \beta_2, \beta_3) = {\rm e}^{2\ii\theta} \left[ {\rm
        e}^{2\ii(\bar\beta - \vp)} 
        + {\rm e}^{2\ii(\bar\beta + \vp)} + {\rm e}^{2\ii(-\bar\beta)} \right]
        \equiv {\rm e}^{2\ii\theta} 
        \bar g,
\end{equation}
we find for the three terms:
\be
g^{-1} {\rm e}^{2\ii(-\vp_1 + \vp_2 + \vp_3)}
         =  \bar g^{-1} {\rm e}^{2\ii (\mu^\prime + \bar \vp - \phi_3)}
         \;;\;\; 
g^{-1} {\rm e}^{2\ii(\vp_1 - \vp_2 + \vp_3)}
         =  \bar g^{-1} {\rm e}^{2\ii (\mu^\prime + \bar \vp + \phi_3)} \;
         ;\;\; 
g^{-1} {\rm e}^{2\ii(\vp_1 + \vp_2 - \vp_3)}
        =  \bar g^{-1} {\rm e}^{2\ii (\mu^\prime - \bar \vp)}. 
\ee
In all three cases, we get the integral $\int_0^{2\pi} \dd \mu^\prime
\exp(2\ii\mu^\prime)
\exp\eck{-\ii A_3 \sin(\mu^\prime - \alpha_3)} = 2\pi \exp(2\ii\alpha_3)
{\rm J}_2(A_3)$.
Finally,
\begin{eqnarray}
& & b(\ell_1, \ell_2, \vp) = \frac{2\pi}{3\bar g} \int_0^\infty \dd x_1\, x_1
        \int_0^\infty \dd 
        x_2 \,x_2 \int_0^{2\pi} \d\phi_3\;
        {\rm e}^{2\ii \alpha_3}\, {\rm J}_2(A_3)\nonumber\\
        & & \qquad\quad \times \left[ {\rm e}^{2\ii(\bar\vp - \phi_3)} \Gamma^{(1)}(x_1, x_2, \phi_3) +
        {\rm e}^{2\ii(\bar\vp + \phi_3)} \Gamma^{(2)}(x_1, x_2, \phi_3) +
        {\rm e}^{-2\ii\bar\vp} \Gamma^{(3)}(x_1, x_2, \phi_3) \right].
\elabel{qsmall1}
\end{eqnarray}

\subsection{\llabel{Comm}Comments}
After having seen the relations of the 3PCF in terms of the bispectrum, one
is not surprised to find that their inversion derived in this section also is
of considerable complexity. This can again be compared to the
case of second-order statistics, where the power spectrum can be written in
terms of the correlation function through an integration. The integration
extends over all angular scales (as is also the case here), and so the direct
inversion will always be of limited accuracy since the correlation functions
can only be measured on a finite range of angular scales. In order to see the
range of application of the previous relations, numerical simulations are
probably required.

The foregoing equations also allow us in principle to express the 3PCF
$\Gamma^{(1)}$ in terms of $\Gamma^{(0)}$, by using the expression
(\ref{eq:Gam1-2}) for $\Gamma^{(1)}$ and substituting the bispectrum in this
equation by $\Gamma^{(0)}$, using (\ref{eq:Q-CF-1}). Whereas the corresponding
equations can be reduced to a three-dimensional integral, they are fairly
complicated; therefore, we shall not reproduce them
here. 

Given the measured correlation functions from a cosmic shear survey, there is
no guarantee that the bispectrum estimates from (\ref{eq:qsmall}) and
(\ref{eq:qsmall1}) will agree, even if we ignore noise and measurement
errors. The two results will agree only if the shear field is derivable from
an underlying convergence field, i.e., if the shear is a pure E-mode
field. Significant differences between the two estimates would then signify
that there is a B-mode contribution to the shear. In the case of
second-order statistics, the separation between E- and B-modes is most
conveniently done in terms of the aperture statistics (see Crittenden et al.\
2002, hereafter CNPT); we shall therefore turn to the aperture measures of the
third-order shear statistics in the next section.

At first sight, it may appear surprising that the expression (\ref{eq:qsmall})
for the bispectrum is always real, even though we have not constrained
$\Gamma^{(0)}$ to correspond to a pure E-mode field (in fact, we would not
really be able to put this constraint on the 3PCF -- compare the 2PCF: only by
combining the two correlation functions $\xi_\pm$ can one separate E- from
B-modes). The only assumption we made was that the shear field is parity
invariant. This can be understood as follows: We can describe a general shear
field by the Fourier transform relation (\ref{eq:FTrel}) if we formally
replace the convergence by $\kappa(\vc X)=\kappa^{\rm E}(\vc X)+{\rm
i}\kappa^{\rm B}(\vc X)$, where $\kappa^{\rm E}$ gives rise to a pure E-mode
shear field, and $\kappa^{\rm B}$ corresponds to a pure B-mode shear
(SvWM). Considering the triple correlator of this complex $\kappa$, one finds
that its real part consists of terms $\ave{(\kappa^{\rm E})^3}$ and
$\ave{\kappa^{\rm E}(\kappa^{\rm B})^2}$, whereas the imaginary part has
contributions $\ave{(\kappa^{\rm E})^2\kappa^{\rm B}}$ and $\ave{(\kappa^{\rm
B})^3}$. As shown by Schneider (2003), the latter two terms are strictly zero
for a parity-invariant shear field, so that the fact that (\ref{eq:qsmall})
is real is fully consistent with the vanishing of the imaginary part of the
triple correlator of the complex $\kappa$ -- both are due to the assumed
parity invariance. This argument then also implies that the resulting
expression (\ref{eq:qsmall}) for $b$ contains both E- and B-modes. 
One can separate E- and B-modes of the bispectrum by suitably combining the
expressions (\ref{eq:qsmall}) and (\ref{eq:qsmall1}). As we shall discuss in 
Sect.~\ref{sc:summary}, the E-mode bispectrum is obtained by 
\be
b^{\rm E}= \eck{b({\rm Eq. }\ref{eq:qsmall})+ 3 b({\rm
Eq. }\ref{eq:qsmall1})} /4\;. 
\elabel{b-mixt}
\ee

\def\com#1{{\breve#1}}
\section{\llabel{Aperture}Aperture statistics}
We have seen that it is possible to calculate the 3PCF in terms of the
bispectrum, and in principle also to invert this relation. However,
the resulting integrals are very cumbersome to evaluate numerically,
owing to the various oscillating factors. It therefore would be useful
to find some statistics that can be easily calculated in terms of the
directly measurable 3PCF, but which can also be easily related to the
bispectrum. In their very interesting paper, JBJ
considered the aperture measures, which have been demonstrated to be
very useful in the case of second-order statistics. The aperture mass
centered on the origin of the coordinate system is defined as
\be
M_{\rm ap}(\theta)=\int \d^2\vt\;U_\theta(|\vc\vt|)\,\kappa(\vc\vt)
=\int \d^2\vt\;Q_\theta(|\vc\vt|)\,\gamma_{\rm t}(\vc\vt)\;,
\elabel{Mapdef}
\ee
where $U_\theta(\vt)$ is a filter function of characteristic radius
$\theta$, the filter function
\be
Q_\theta(\vt)={2\over\vt^2}\int_0^\vt\d\vt'\,\vt'\,U_\theta(\vt')
-U_\theta(\vt) 
\elabel{Qdef}
\ee
is related to $U_\theta(\vt)$, and the second equality in (\ref{eq:Mapdef}) is
true as long as $U_\theta$ is a compensated filter,
i.e. $\int\vt\,\vt\,U_\theta(\vt)=0$, as has been shown by Kaiser et al.\
(1994) and Schneider (1996). $\gamma_{\rm t}$ is the shear component tangent
to the center of the aperture, i.e., the origin. Hence, $\gamma_{\rm
t}(\vc\vt)+{\rm i}\gamma_\times(\vc\vt) =-\gamma\, \com\vt^{*2}/|\vc\vt|^2$,
where here and in the following we use the notation that a vector $\vc
x=(x_1,x_2)$ can also be represented by a complex number $\com{x}=x_1+{\rm
i}x_2$. Hence, $\com\vt^{*2}/|\vc\vt|^2$ is nothing but the phase factor ${\rm
e}^{-2{\rm i}\phi}$, where $\phi$ is the polar angle of $\vc\vt$.

The aperture mass as a statistics for cosmic shear was introduced by
SvWJK who showed that the dispersion $\ave{M_{\rm
ap}^2(\theta)}$ of the aperture mass is given as the integral over the
power spectrum of the projected mass density $\kappa$, convolved with
a filter function which is the square of the Fourier transform of
$U_\theta$. SvWJK derived this filter function for a family of
functions $U_\theta$ which have a finite support. These filter
functions turn out to be quite narrow, so that $\ave{M_{\rm
ap}^2(\theta)}$ provides very localized information about the power
spectrum (see also Bartelmann \& Schneider 1999). Furthermore, SvWJK
calculated the skewness of $M_{\rm ap}(\theta)$ in the frame of
second-order perturbation theory for the growth of structure. As it
turned out, the resulting equations are quite cumbersome, which is in
part related to the fact that the Fourier transform of the functions
$U_\theta$ chosen contains a Bessel function.

CNPT suggested an alternative form of the function
$U_\theta$. When we write $U_\theta(\vt)=\theta^{-2}\,u(\vt/\theta)$, then the
filter used by CNPT is
\be
u(x)={1\over 2\pi}\rund{1-{x^2\over 2}}\,{\rm e}^{-x^2/2}\; ;\;\;
\hat u(\eta)=\int \d^2 x\;u(|\vc x|)\,{\rm e}^{{\rm i}\vc\eta\cdot\vc x}
={\eta^2\over 2}\,{\rm e}^{-\eta^2/2}\; ; \;\;
Q_\theta(\vt)={\vt^2\over 4\pi\theta^4}\exp\rund{-{\vt^2\over 2\theta^2}}\;.
\elabel{u+uFour}
\ee
Hence, this filter function does not have finite support; this is, however,
only a small disadvantage for employing it since it cuts off very quickly for
distances larger than a few $\theta$. This disadvantage is more than
compensated by the convenient analytic properties of this filter. 

A further advantage of using aperture measures is that $M_{\rm ap}$, as
calculated from the rightmost expression in (\ref{eq:Mapdef}), is sensitive
only to an E-mode shear field (see CNPT and SvWM for a discussion of the
E/B-mode decomposition of shear fields). Hence, 
defining the complex number
\be
M(\theta):=M_{\rm ap}(\theta)+{\rm i}M_\perp(\theta)
=\int \d^2\vt\;Q_\theta(|\vc\vt|)\eck{\gamma_{\rm t}(\vc\vt)+{\rm
i}\gamma_\times(\vc\vt)}=-\int \d^2\vt\;Q_\theta(|\vc\vt|)
\gamma(\vc\vt) \,\com{\vt}^{*2}/|\vc\vt|^2 \;,
\elabel{Mdef}
\ee 
$M_{\rm ap}(\theta)$ vanishes identically for B-modes, whereas
$M_\perp(\theta)$ yields zero for a pure E-mode field. Thus, the
aperture measures are ideally suited to separating E- and B-modes of
the shear.

CNPT and SvWM have shown that the dispersions $\ave{M_{\rm
ap}^2(\theta)}$ and $\ave{M_\perp^2(\theta)}$ can be expressed as an
integral over the two-point correlation functions of the shear. Since
the correlation functions are the best measured statistics on real
data (as they are insensitive to the gaps and holes in the data
field), this property allows an easy calculation of the aperture
dispersions from the data. JBJ showed that the third-order moments of
the aperture measures can likewise be expressed by the shear 3PCF, and
they derived the corresponding relations explicitly -- they are
remarkably simple. The fact that such explicit results can be obtained
is tightly related to the choice of the filter function
(\ref{eq:u+uFour}); for a filter function with strictly finite
support, the resulting expressions are very messy (indeed, we have
derived such an expression for the filter function used in SvWJK, but
it is so complicated that it will most likely be useless for any
practical work).

\subsection{An alternative derivation of the third-order aperture mass}
We shall here rederive one of the results from JBJ making
use of the results obtained in Sect.~\ref{sc:Bispect}; the agreement
of the resulting expression with that of JBJ provides a convenient
check for the correctness of the results in Sect.~\ref{sc:Bispect}.
In a first step, we
express $\ave{M_{\rm ap}^3(\theta)}$ in terms of the bispectrum. Using the
first definition in (\ref{eq:Mapdef}), we find that
\bea
\ave{M_{\rm ap}^3(\theta)}&=&
\int\d^2\vt_1\,U_\theta(|\vc\vt_1|)
\int\d^2\vt_2\,U_\theta(|\vc\vt_2|)
\int\d^2\vt_3\,U_\theta(|\vc\vt_3|) \nonumber \\ &\times&
\int{\d^2\ell_1\over (2\pi)^2}
\int{\d^2\ell_2\over (2\pi)^2}
\int{\d^2\ell_3\over (2\pi)^2}
{\rm e}^{-{\rm i}(\vc\ell_1\cdot \vc\vt_1 + \vc\ell_2\cdot \vc\vt_2 +
\vc\ell_3\cdot \vc\vt_3)}
\ave{\hat\kappa(\vc\ell_1)\hat\kappa(\vc\ell_2)\hat\kappa(\vc\ell_3)} \;.
\elabel{Map3-1}
\eea
When carrying out the $\vt_i$-integrations, the Fourier transforms $\hat
U_\theta$ are obtained. Inserting the bispectrum in the form
(\ref{eq:bispec}) and integrating out the corresponding delta function, one
obtains three identical terms. With $\hat U_\theta(\ell)= \hat u(\theta\ell)$
one finds
\bea
\ave{M_{\rm ap}^3(\theta)}&=&3\int{\d^2\ell_1\over (2\pi)^2}
\int{\d^2\ell_2\over (2\pi)^2}\,B(\vc\ell_1,\vc\ell_2)\,
\hat u(\theta|\vc\ell_1|)\,\hat u(\theta|\vc\ell_2|)\,
\hat u(\theta |\vc\ell_1+\vc\ell_2|)
\elabel{Map3-2} \\
&=&{3\over (2\pi)^3}
\int\d\ell_1\,\ell_1 \int\d\ell_2\,\ell_2
\int\d\vp\;b(\ell_1,\ell_2,\vp)\;
\hat u(\theta\ell_1)\;\hat u(\theta\ell_2)\;
\hat u\rund{\theta\sqrt{\ell_1^2+\ell_2^2+2\ell_1\ell_2\cos\vp}}\;.
\elabel{Map3-3}
\eea
We shall discuss this result in the next subsection; here, we want to use
(\ref{eq:Map3-2}) and obtain an explicit equation for $\ave{M_{\rm ap}^3}$ in
terms of the 3PCF of the shear. For better comparison with JBJ,
we shall slightly change our notation for the 3PCF. Up to now we have
labeled the sides of the triangle formed by the three points $\vc X_i$ by the
vectors $\vc x_i$ as defined in Sect.~\ref{sc:preli}. The corresponding
natural components of the 3PCF were then denoted by
$\Gamma^{(n)}(x_1,x_2,x_3)=\Gamma^{(n)}(\vc x_1,\vc x_2)$. We shall now define
the three points $\vc X_i$ in the form $\vc X_1=\vc X_3+\vc y_1$, $\vc X_2=\vc
X_3 + \vc y_2$, and then define
\be
\ave{\gamma^{(\rm x)}(\vc X_1)\,\gamma^{(\rm x)}(\vc X_2)
\,\gamma^{(\rm x)}(\vc X_3)} = \tilde\Gamma_{\rm x}^{(0)}(\vc y_1,\vc y_2) \;
; \;\;
\ave{\gamma^{(\rm x)}(\vc X_1)\,\gamma^{(\rm x)}(\vc X_2)
\,\gamma^{(\rm x)*}(\vc X_3)} = \tilde\Gamma_{\rm x}^{(3)}(\vc y_1,\vc y_2) \;,
\elabel{Map3-4}
\ee
where the `x' denotes an arbitrary projection of the shear components,
i.e. relative to an arbitrary choice of reference directions. The relation
between the $\Gamma$ and the $\tilde\Gamma$ follows simply from the
definitions of the separation vectors between the points $\vc X_i$, which is
$\vc y_1=\vc x_2$, $\vc y_2=-\vc x_1$, so that
\be
\tilde \Gamma_{\rm x}^{(0)}(\vc y_1,\vc y_2)
=\tilde\Gamma_{\rm x}^{(0)}(\vc x_2,-\vc x_1)
=\Gamma^{(0)}_{\rm x}(\vc x_1,\vc x_2)\;,
\elabel{GamTrans}
\ee
and analogously for the other components of the 3PCF. 
Now, from combining (\ref{eq:Q-CF-1}) with (\ref{eq:Map3-2}) and using our new
notation for the 3PCF, one finds
\be
\ave{M_{\rm ap}^3(\theta)}=\int\d^2 y_1\int\d^2 y_2\;
\tilde\Gamma^{(0)}_{\rm cart}(\vc y_1,\vc y_2)
\int {\d^2\ell_1\over (2\pi)^2} \int {\d^2\ell_2\over (2\pi)^2} \;
\hat u(\theta|\vc\ell_1|)\,\hat u(\theta|\vc\ell_2|)\,
\hat u(\theta |\vc\ell_1+\vc\ell_2|) \,
{\rm e}^{-2{\rm i}\sum\beta_i}\,
{\rm e}^{{\rm i}(\vc\ell_1\cdot\vc y_1+\vc\ell_2\cdot\vc y_2)} \;,
\elabel{Map3-5}
\ee
where the subscript `cart' denotes the Cartesian components of the 3PCF.
Inserting the Fourier transforms of $u$ from (\ref{eq:u+uFour}) and using  
$\hat u(\theta|\vc\ell_i|)\,{\rm e}^{-2{\rm i}\beta_i}
=\hat u(\theta|\vc\ell_i|) \com{\ell_i}^{*2}/|\com{\ell_i}|^2
=(\theta^2\com{\ell_i}^{*2} /2)\,{\rm
e}^{-|\com{\ell_i}|^2\theta^2/2}$ (where we used, as before, the notation
$\com\ell=\ell_1+{\rm i}\ell_2$),
(\ref{eq:Map3-5}) becomes
\bea
\ave{M_{\rm ap}^3(\theta)}&=&\int\d^2 y_1\int\d^2 y_2\;
\tilde\Gamma^{(0)}_{\rm cart}(\vc y_1,\vc y_2)
\int {\d^2\ell_1\over (2\pi)^2} \int {\d^2\ell_2\over (2\pi)^2} \;
\com{\ell_1}^{*2}\,\com{\ell_2}^{*2}\,(\com{\ell_1}^*+\com{\ell_2}^*)^2
\nonumber \\ &\times&
\exp\eck{-{\rund{|\com{\ell_1}|^2+|\com{\ell_2}|^2+|\com{\ell_1}
+\com{\ell_2}|^2}\theta^2 \over 2} +{\rm i}\rund{\vc y_1\cdot\vc\ell_1+\vc
y_2\cdot \vc\ell_2}} \;.
\elabel{Map3-6}
\eea
The $\ell_i$-integrations can now be carried out, by noting that the
exponential 
is just a quadratic function of the integration variables, and it is
multiplied by a polynomial. The straightforward, but tedious calculation has
been carried out with Mathematica (Wolfram 1999); the result of the
integration then depends on the $\vc y_i$. Substituting the $\vc y_i$ in
favor of the vectors
\be
\vc q_1={2 \vc y_1-\vc y_2\over 3}\; ;\;\;
\vc q_2={2 \vc y_2-\vc y_1\over 3}\; ;\;\;
\vc q_3=-{\vc y_1+\vc y_2\over 3}\;,
\elabel{qvecs}
\ee
which are the vectors connecting the center of mass of the triangle with its
three corners, one can put the result in the form
\be
\ave{M_{\rm ap}^3(\theta)}=-{1\over 24(2\pi)^2}\int{\d^2 y_1\over
\theta^2}\int{\d^2 y_2\over \theta^2} \;
\tilde\Gamma^{(0)}_{\rm cart}(\vc y_1,\vc y_2)
\,{\com{q_1}^{*2}\,\com{q_2}^{*2}\,\com{q_3}^{*2}\over \theta^6}\,
\exp\rund{-{|\com{q_1}|^2+|\com{q_2}|^2+|\com{q_3}|^2
\over 2 \theta^2}} \;.
\elabel{Map3-7}
\ee
Employing now the transformation laws of the natural components of the 3PCF as
derived in Paper~I, one sees that the squares of the
complex conjugates of the $q_i$ can be used to obtain the 3PCF with the shear
projected along the direction towards the center of mass of the triangle,
i.e., 
\be
\tilde\Gamma^{(0)}_{\rm cart}(\vc y_1,\vc y_2)
\com{q_1}^{*2}\,\com{q_2}^{*2}\,\com{q_3}^{*2}
=-\tilde\Gamma^{(0)}_{\rm cen}(\vc y_1,\vc y_2) \,
|\com{q_1}|^2\,|\com{q_2}|^2\,|\com{q_3}|^2 \;.
\elabel{Map3-8}
\ee
After this projection of the 3PCF, the integrand depends only on the absolute
values of the $\vc y_1$ and the angle $\psi$ between them. By carrying out one
more integration, one finally obtains
\be
\ave{M_{\rm ap}^3(\theta)}={1\over 24}\int{\d y_1\,y_1\over
\theta^2}\int{\d y_2\,y_2\over \theta^2} \int{\d\psi\over (2\pi)} \;
\tilde\Gamma^{(0)}_{\rm cen}(y_1,y_2,\psi)\,
{|\com{q_1}|^2\,|\com{q_2}|^2\,|\com{q_3}|^2 \over \theta^6}\,
\exp\rund{-{|\com{q_1}|^2+|\com{q_2}|^2+|\com{q_3}|^2
\over 2 \theta^2}} \;,
\elabel{Map3-9}
\ee
with
\be
|\com{q_1}|^2={4y_1^2-4y_1y_2\cos\psi+y_2^2\over 9} \; ; \;\;
|\com{q_2}|^2={y_1^2-4y_1y_2\cos\psi+4y_2^2\over 9} \; ; \;\;
|\com{q_3}|^2={y_1^2+2y_1y_2\cos\psi+y_2^2\over 9} \;,
\ee
and thus
\be
|\com{q_1}|^2+|\com{q_2}|^2+|\com{q_3}|^2={2\over 3}\;
\rund{y_1^2+y_2^2-y_1y_2\cos\psi} \;.
\ee
The result (\ref{eq:Map3-9}) agrees with equation (44) of JBJ, after the
projection of the 3PCF has been accounted for. It should be noted that in the
case considered here, where the 3PCF was explicitly obtained in terms of the
bispectrum of the convergence which, owing to parity invariance, is real
(cf. the discussion in Sect.~\ref{sc:Comm}), the expression
(\ref{eq:Map3-3}) is real, and hence (\ref{eq:Map3-9}) also is real in this
case. This can be seen explicitly, since the value of 
$\tilde\Gamma^{(0)}_{\rm cen}$ at $\psi$ is just the complex conjugate
one of that at $-\psi$. 

We like to point out that this derivation has shown two interesting
aspects: first, the natural component $\Gamma^{(0)}$ of the shear 3PCF
arises naturally in this context, confirming the hypothesis of Paper\
I that this combination of components of the correlation functions is
indeed useful. Second, the derivation shows that the projection of the
shear onto the centroid is the most convenient projection in this
particular application.

\subsection{\llabel{Gene}Generalization: Third-order aperture statistics with
different filter radii}
How important is the third-order aperture statistics for investigating the
third-order statistical properties of the cosmic shear? In order to discuss
this question, we shall first consider the analogous situation for the
second-order statistics. There, as mentioned before, the aperture mass
dispersion is a filtered version of the power spectrum $P_\kappa(\ell)$ of the
underlying convergence; for the function $U_\theta$ considered here, one has
\be
\ave{M_{\rm ap}^2(\theta)}=\int{\d\ell\,\ell\over (2\pi)}\;
P_\kappa(\ell)\,{\theta^4\,\ell^4\over 4}\,{\rm e}^{-\theta^2\ell^2}\;;
\elabel{Map2-1}
\ee
hence, the filter function relating $P_\kappa(\ell)$ and $\ave{M_{\rm
ap}^2(\theta)}$ 
is very narrow, and unless the power spectrum
exhibits sharp features, the function $\ave{M_{\rm ap}^2(\theta)}$ contains
basically all the information available for second-order shear statistics (not
quite -- see below).
The analogous equation to (\ref{eq:Map2-1}) for third-order statistics is
given in (\ref{eq:Map3-3}). The function $\hat u$ is very narrowly peaked at
around $\ell\theta\sim 1$, and there is one factor of $\hat u$ for each of the
three sides of a triangle in $\ell$-space. This implies that in the
integration of (\ref{eq:Map3-3}) the bispectrum is probed only in regions of
$\ell$-space where $\ell_1\sim \ell_2\sim
|\vc\ell_1+\vc\ell_2|\sim 1/\theta$. Thus, $\ave{M_{\rm ap}^3(\theta)}$ probes
the bispectrum essentially only for equilateral triangles in Fourier
space. For this reason, the function  $\ave{M_{\rm ap}^3(\theta)}$ cannot carry
the full information of the bispectrum; it merely yields part of this
information. 

On the other hand, (\ref{eq:Map3-3}) immediately suggests how to improve 
this situation: if we define the aperture mass statistics with three different
filter radii $\theta_i$, we can probe the bispectrum at wavevectors whose
lengths are $\ell_i\sim 1/\theta_i$, and by covering a wide range of
$\theta_i$, one can essentially probe the bispectrum over the full
$\ell$-space. Indeed,
\bea
\ave{M_{\rm ap}(\theta_1)M_{\rm ap}(\theta_2)M_{\rm ap}(\theta_3)}
&=& \int{\d^2\ell_1\over (2\pi)^2}
\int{\d^2\ell_2\over (2\pi)^2}\,B(\vc\ell_1,\vc\ell_2)\,
\Big[ \, \hat u(\theta_1|\vc\ell_1|)\,\hat u(\theta_2|\vc\ell_2|)\,
\hat u(\theta_3 |\vc\ell_1+\vc\ell_2|) \Big. \nonumber \\
& & \Big. + \hat u(\theta_2 |\vc\ell_1|) \, \hat u(\theta_3 | \vc \ell_2 |)
\, \hat u(\theta_1 |\vc \ell_1 + \vc \ell_2|) + \hat u(\theta_3 |\vc\ell_1|)
\,\hat u(\theta_1 | \vc \ell_2 |) \, \hat u(\theta_2 |\vc \ell_1 +
\vc \ell_2|) \, \Big] \\
= {1\over (2\pi)^3}\int\d\ell_1  \ell_1 \int\d\ell_2 \!\!& \ell_2 &\!\!
\int\d\vp\;b(\ell_1,\ell_2,\vp)\eck{
\hat u(\theta_1 \ell_1)\,\hat u(\theta_2 \ell_2)\,
\hat
u\rund{\theta_3\sqrt{\ell_1^2+\ell_2^2+2\ell_1\ell_2\cos\vp}}+\hbox{2
terms}} \;, \nonumber
\eea
which illustrates what was said above. Thus, this third-order statistics is
expected to be as important for the third-order shear statistics as is the
aperture mass dispersion for second-order shear statistics. The fact that we
do not gain additional information by considering different filter scales for
the second-order $M_{\rm ap}$ statistics follows from the fact that
\be
\ave{M_{\rm ap}(\theta_1)M_{\rm ap}(\theta_2)}
= \int{\d\ell\,\ell\over (2\pi)}\;
P_\kappa(\ell)\,{\theta_1^2\theta_2^2\,\ell^4\over 4}\,{\rm e}^{-(\theta_1^2+
\theta_2^2)\ell^2/2} 
={4\theta_1^2\theta_2^2\over \rund{\theta_1^2+\theta_2^2}^2}
\ave{M_{\rm ap}^2\rund{\sqrt{\theta_1^2+\theta_2^2\over 2}}}\;.
\ee
Whereas the fact that the mixed correlator can be expressed exactly in terms
of the dispersion at an average angle depends on the special filter function
considered here, it nevertheless shows that one does not gain additional
information when considering the covariance of $M_{\rm ap}$.

We shall now calculate the triple correlator of $M_{\rm ap}$ for three
different filter radii in terms of the shear 3PCF, essentially using the same
method as JBJ. For that, we first calculate the
third-order statistics of the complex aperture measure $M$, as defined in
(\ref{eq:Mdef}):
\bea
\ave{M(\theta_1)M(\theta_2)M(\theta_3)}
&\equiv&\ave{M^3}(\theta_1,\theta_2,\theta_3)=
-\int\d^2 X_1 \int\d^2 X_2 \int\d^2 X_3\;
Q_{\theta_1}(|\vc X_1|) Q_{\theta_2}(|\vc X_2|) Q_{\theta_3}(|\vc X_3|)
\nonumber \\ &\times&
\ave{\gamma(\vc X_1)\gamma(\vc X_2)\gamma(\vc X_3)}\,{\rm e}^{-2{\rm
i}(\phi_1+\phi_2+\phi_3)}\;, 
\elabel{Map3-13}
\eea
where the $\phi_i$ are the polar angles of the vectors $\vc X_i$. Writing, as
before, $\vc X_1=\vc X_3+\vc y_1$, $\vc X_2=\vc X_3+\vc y_2$, replacing the
phase factors by ${\rm e}^{-2{\rm i}\phi_i}= \com{X_i}^{*2}/|\com{X_i}|^2$,
and inserting the definitions of the $Q_\theta$, one obtains
\bea
\ave{M^3}(\theta_1,\theta_2,\theta_3)
&=&{- 1\over (4\pi)^3\theta_1^4\theta_2^4\theta_3^4}
\int\d^2 y_1\int\d^2 y_2\;\tilde\Gamma^{(0)}_{\rm cart}(\vc y_1,\vc y_2)
\nonumber \\ &\times&
\int\d^2 Y\,\com{Y}^{*2}(\com{Y}^*+\com{y_1}^*)^2 (\com{Y}^*+\com{y_2}^*)^2
\exp\eck{-\rund{ {|\vc Y+\vc y_1|^2\over 2 \theta_1^2} + {|\vc Y+\vc
y_2|^2\over 2 \theta_2^2} +    {|\vc Y|^2\over 2 \theta_3^2}} } \;,
\elabel{Map3-14}
\eea
where we set for ease of notation the dummy variable $\vc X_3\equiv \vc Y$. 
The $Y$-integration is again over the exponential of a second-order polynomial
in the integration variable, times a polynomial, and thus straightforward to
integrate, but tedious. Employing Mathematica does most of the job, though
its output needed to be further simplified. The result is
\be
\ave{M^3}(\theta_1,\theta_2,\theta_3)
={S\over 24}\int{\d y_1\,y_1\over \Theta^2} \int{\d y_2\,y_2\over \Theta^2}
\int_0^{2\pi}{\d\psi\over (2\pi)}\,\tilde\Gamma^{(0)}_{\rm cen}(y_1,y_2,\psi) 
{|\com{q_1}|^2\,|\com{q_2}|^2\,|\com{q_3}|^2 \over \Theta^6}\,
f_1^{*2}\,f_2^{*2}\,f_3^{*2}\,
{\rm e}^{-Z} \;,
\elabel{M3}
\ee
where
\be
\Theta^2=\sqrt{\theta_1^2\theta_2^2 +
\theta_1^2\theta_3^2+\theta_2^2\theta_3^2\over 3} \; ; \;\;
S={\theta_1^2\theta_2^2\theta_3^2\over \Theta^6} \;,
\ee
\be
Z={(-\theta_1^2+2\theta_2^2+2\theta_3^2)|\com{q_1}|^2
+(2\theta_1^2-\theta_2^2+2\theta_3^2)|\com{q_2}|^2
+(2\theta_1^2+2\theta_2^2-\theta_3^2)|\com{q_3}|^2
\over
6 \Theta^4 } \;,
\elabel{Map3-15}
\ee
\be
f_1={\theta_2^2+\theta_3^2\over 2\Theta^2} +
{(\com{q_2}-\com{q_3})\com{q_1}^*\over |\com{q_1}|^2}
\,{\theta_2^2-\theta_3^2\over 6\, \Theta^2} \;,\; \;\;
f_2={\theta_1^2+\theta_3^2\over 2\Theta^2} +
{(\com{q_3}-\com{q_1})\com{q_2}^*\over |\com{q_2}|^2}
\,{\theta_3^2-\theta_1^2\over 6\, \Theta^2} \;,\; \;\;
f_3={\theta_1^2+\theta_2^2\over 2\Theta^2} +
{(\com{q_1}-\com{q_2})\com{q_3}^*\over |\com{q_3}|^2}
\,{\theta_1^2-\theta_2^2\over 6\, \Theta^2} \;.
\elabel{Map3-16}
\ee
The choice of the various quantities defined above was made such that $S$,
$Z$, and the $f_i$ are dimensionless, and that they become very simple
if all $\theta_i$ are equal. Consider this special case next, i.e., let
$\theta_1=\theta_2=\theta_3=\theta$. Then, $\Theta=\theta$, $S=1$,
$f_1=f_2=f_3=1$, and
$Z=\rund{|\com{q_1}|^2+|\com{q_2}|^2+|\com{q_3}|^2}/(2\theta^2)$. 
Thus, we recover the result (\ref{eq:Map3-9}) in this case which was shown
above to agree with the result from JBJ. The difference
between (\ref{eq:Map3-9}) and (\ref{eq:M3}) is that the former has been
derived in this paper from the bispectrum of the convergence, and therefore is
strictly real, whereas (\ref{eq:M3}) has been calculated directly in
terms of the shear 3PCF and thus applies to arbitrary shear fields, containing
both E- and B-modes. For reference, we explicitly give the combinations of the
$\com{q_i}$ appearing in the $f_i$ above,
\[
{(\com{q_2}-\com{q_3})\com{q_1}^*\over |\com{q_1}|^2}
={3y_2(2y_1\,{\rm e}^{{\rm i}\psi}-y_2)\over 4
y_1^2-4y_1y_2\cos\psi+y_2^2}
\; ;\;
{(\com{q_3}-\com{q_1})\com{q_2}^*\over |\com{q_2}|^2}
={3y_1(y_1-2y_2\,{\rm e}^{-{\rm i}\psi})\over
y_1^2-4y_1y_2\cos\psi+4y_2^2}
\; ;\;
{(\com{q_1}-\com{q_2})\com{q_3}^*\over |\com{q_3}|^2}
=3\,{y_2^2-y_1^2+2{\rm i}y_1y_2\sin\psi\over
y_1^2+2y_1y_2\cos\psi+y_2^2} \,.
\]
One expects that $\ave{M^3}$ is symmetric with respect to any
permutation of its arguments. Indeed, one can show explicitly that
(\ref{eq:M3}) is symmetric with respect to interchanging $\theta_1$
and $\theta_2$. Performing this interchange, changing the variables
of integration as $y_1 \to y_2$, $y_2\to y_1$, $\psi\to -\psi$, and
making use of the fact that $\tilde\Gamma^{(0)}_{\rm
cen}(y_1,y_2,\psi)=\tilde\Gamma^{(0)}_{\rm cen}(y_2,y_1,2\pi-\psi)$,
one finds that these transformations lead to $f_1\to f_2$,
$f_2\to f_1$, $f_3\to f_3$, and $Z$ is unchanged. To show the symmetry
with respect to even permutations of the arguments, one needs to employ
the symmetry of $\Gamma^{(0)}(x_1,x_2,x_3)=\Gamma^{(0)}(x_2,x_3,x_1)$,
and then use either $X_1$ or $X_2$ as the reference point in the
derivation. This then leads to a cyclic permutation of the $q_i$ and
the $f_i$, and thus leaves (\ref{eq:M3}) invariant.

\begin{figure}
\resizebox{\hsize}{!}{
\includegraphics{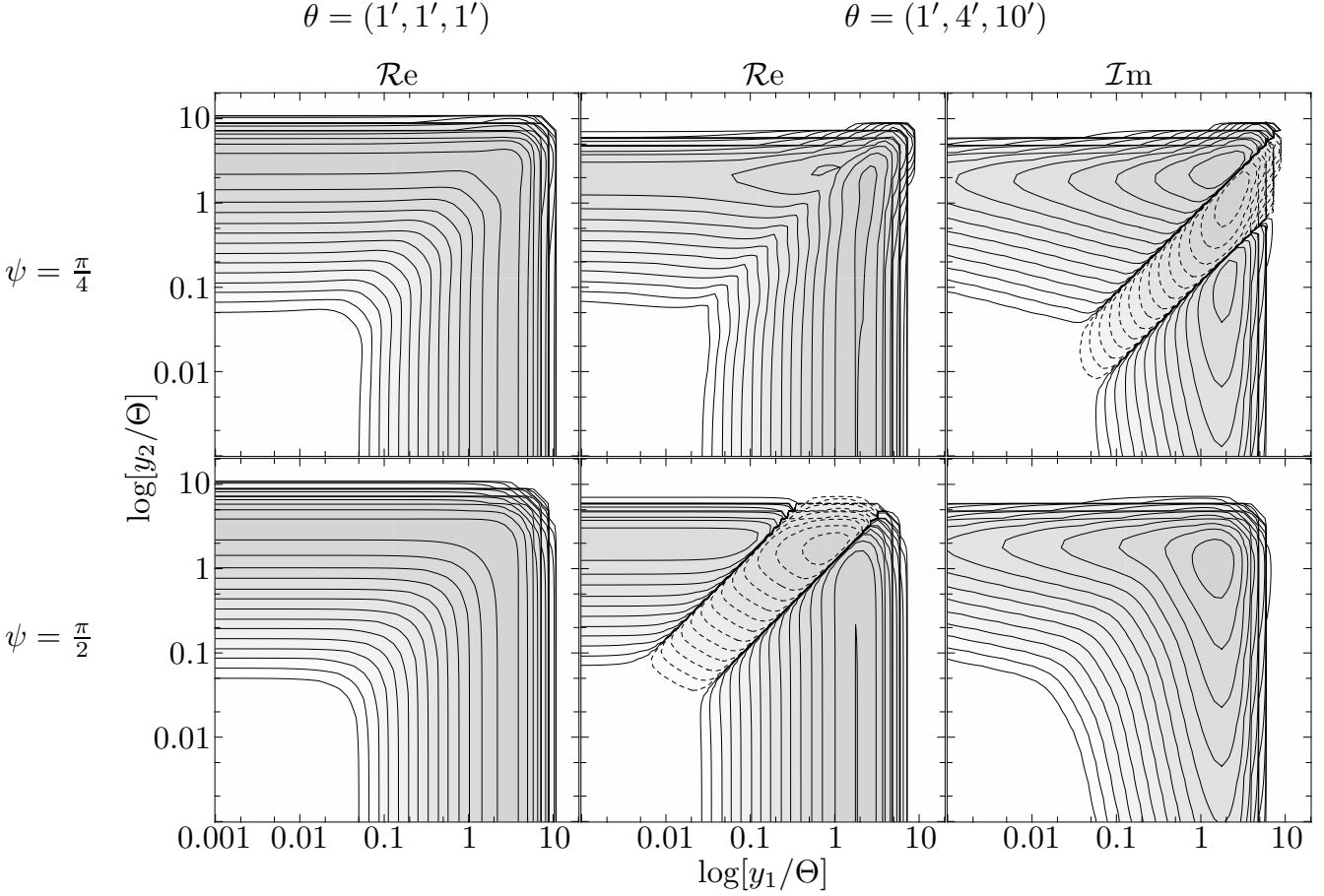}
\bigskip
}
\caption{Contours of the integration function ${|\com{q_1}|^2\,|\com{q_2}|^2\,|\com{q_3}|^2 \over \Theta^6}\,
f_1^{*2}\,f_2^{*2}\,f_3^{*2}\, {\rm e}^{-Z}$ -- see (\ref{eq:M3}) -- as a
function of $y_1$ and $y_2$ for fixed $\psi$ (upper row: $\psi=\pi/4$,
lower row: $\psi=\pi/2$).
The left-most of the three columns represents the case where all
three aperture radii are equal. The function scales with the aperture
radius. Note that the imaginary part vanishes here because of symmetry.
The two right columns show the real and imaginary part of the
integrand for three different filter radii.
The contour lines are logarithmically spaced with a factor of 5
between successive lines,
starting with $10^{-10}$. Dashed lines correspond to negative values.}
\label{fig2}
\end{figure}

In Fig.\ \ref{fig2} we show the latter part of the integrand in
(\ref{eq:M3}) for the case of three equal apertures
($\theta_1=\theta_2=\theta_3$) and for different aperture sizes. Its
zeros, if any, are lines of constant $y_2/y_1$,
because the function only depends on the ratio of $y_2$ and $y_1$.

\begin{figure}
\resizebox{\hsize}{!}{
\includegraphics{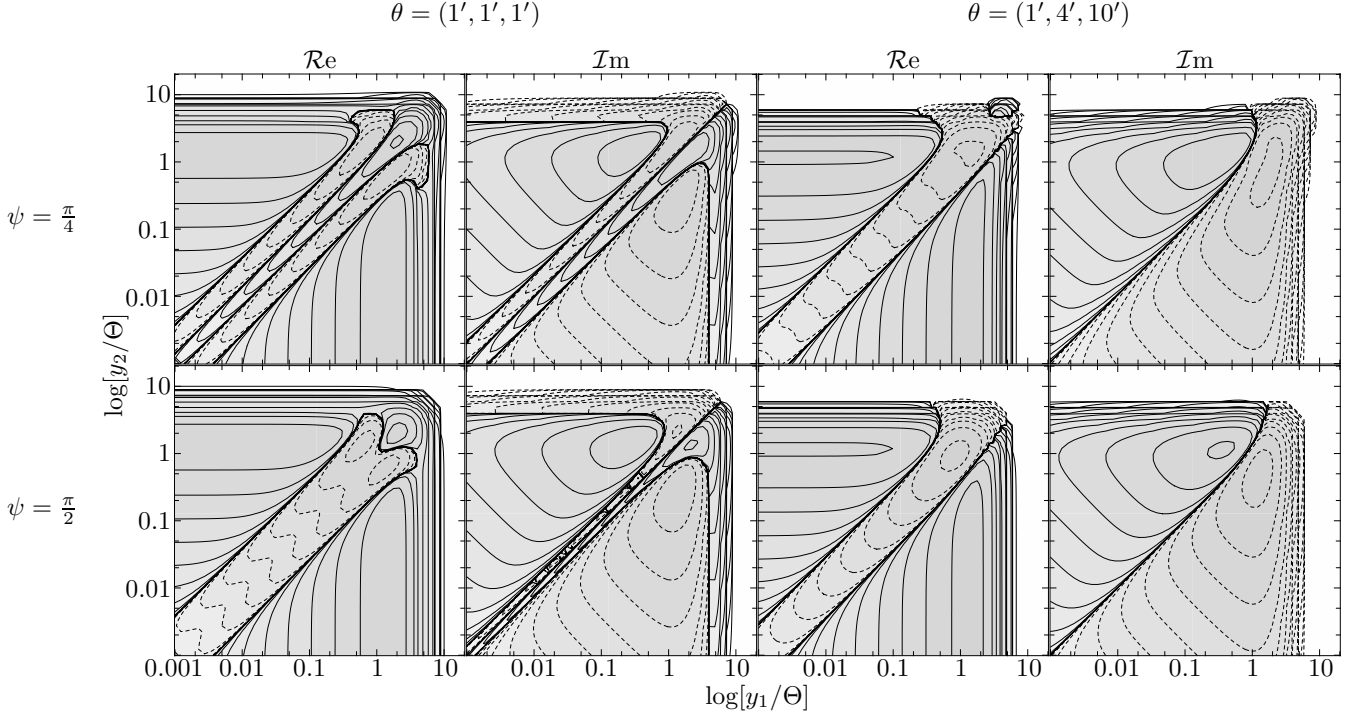}
\bigskip
}
\caption{Contours of the latter part of the integrand in
(\ref{eq:Map3-22}). The plotted function 
is ${\rm e}^{-Z}$ times the term in square brackets. The upper and
lower row correspond to fixed values of $\psi=\pi/4$ and $\pi/2$
respectively. In the left two columns, the real and imaginary part of
the function is shown for the three aperture radii being equal. The
right two panels correspond to three different radii. The contours are
the same as in Fig.\ \ref{fig2}.
}
\label{fig3}
\end{figure}

We next consider the combination of aperture measures
\bea
\ave{M(\theta_1)M(\theta_2)M^*(\theta_3)}
&\equiv&\ave{M^2 M^*}(\theta_1,\theta_2;\theta_3)=
-\int\d^2 X_1 \int\d^2 X_2 \int\d^2 X_3\;
Q_{\theta_1}(|\vc X_1|) Q_{\theta_2}(|\vc X_2|) Q_{\theta_3}(|\vc X_3|)
\nonumber \\ &\times&
\ave{\gamma(\vc X_1)\gamma(\vc X_2)\gamma^*(\vc X_3)}\,{\rm e}^{-2{\rm
i}(\phi_1+\phi_2-\phi_3)}\;, 
\elabel{Map3-17}
\eea
where the semicolon in the arguments of $\ave{M^2 M^*}$ indicates that
this expression is symmetric with respect to interchanging the first
two arguments, but not the third one, of course.
Using the same conventions for labeling the vertices $\vc X_i$ of the
triangle as before, we obtain
\bea
\ave{M^2 M^*}(\theta_1,\theta_2;\theta_3)
&=&{- 1\over (4\pi)^3\theta_1^4\theta_2^4\theta_3^4}
\int\d^2 y_1\int\d^2 y_2\;\tilde\Gamma^{(3)}_{\rm cart}(\vc y_1,\vc y_2)
\nonumber \\ &\times&
\int\d^2 Y\,\com{Y}^{2}(\com{Y}^*+\com{y_1}^*)^2 (\com{Y}^*+\com{y_2}^*)^2
\exp\eck{-\rund{ {|\vc Y+\vc y_1|^2\over 2 \theta_1^2} + {|\vc Y+\vc
y_2|^2\over 2 \theta_2^2} +    {|\vc Y|^2\over 2 \theta_3^2}} } \;,
\elabel{Map3-18}
\eea
After performing the $Y$-integration and a few manipulations to express the
$\vc y_i$ in terms of the $\vc q_i$, we obtain
\bea
\ave{M^2 M^*}(\theta_1,\theta_2;\theta_3)
&=& -{S\over (2\pi)^2}\int{\d^2 y_1\over \Theta^2}
\int{\d^2 y_2\over \Theta^2}\;\tilde\Gamma^{(3)}_{\rm cart}(\vc y_1,\vc y_2)
\;{\rm e}^{-Z}
\nonumber \\ 
&\times&
\Biggl[ {1\over 24}\,{\com{q_1}^{*2}\com{q_2}^{*2}\com{q_3}^{2} \over
\Theta^6} \,f_1^{*2}f_2^{*2} f_3^{2}
-{1\over 9}\,{\com{q_1}^* \com{q_2}^* |\com{q_3}|^2\over \Theta^4}\,
f_1^*f_2^* f_3\, g_3^*
+{1\over 27} \rund{ {\com{q_3}^{*2}\over \Theta^2}\,g_3^{*2}
+{2\theta_1^2\theta_2^2\over \Theta^4}\,
{\com{q_1}^* \com{q_2}^*\over \Theta^2} \,f_1^*f_2^*} \Biggr]\;,
\elabel{Map3-19}
\eea
where we have defined
\be
g_3={\theta_1^2\theta_2^2\over \Theta^4}+
{(\com{q_1}-\com{q_2})\com{q_3}^*\over |\com{q_3}|^2}\,
{\theta_3^2(\theta_2^2-\theta_1^2) \over 3\,\Theta^4} \;,
\elabel{Map3-20}
\ee
and the $f_i$ are as before. This form of the equation is easily compared with
the result obtained by JBJ, by setting
$\theta_1=\theta_2=\theta_3=\theta$, so that 
$\Theta=\theta$, $S=1$,
$f_1=f_2=f_3=g_3=1$, and
$Z=\rund{|\com{q_1}|^2+|\com{q_2}|^2+|\com{q_3}|^2}^2/2$. This then reproduces
their Eq.~(49), except for a different labeling of the $q_i$ (we considered
the complex conjugate shear at the point $\vc X_3$, whereas JBJ did
this at $\vc X_1$). 

We now employ again the relation between the natural components of the shear
3PCF in the Cartesian reference frame and those measured relative to the
center of mass of the triangle (see Paper~I),
\be
\tilde\Gamma^{(3)}_{\rm cart}(\vc y_1,\vc y_2)
\com{q_1}^{*2}\,\com{q_2}^{*2}\,\com{q_3}^{2}
=-\tilde\Gamma^{(3)}_{\rm cen}(\vc y_1,\vc y_2)\,
|\com{q_1}|^2\,|\com{q_2}|^2\,|\com{q_3}|^2 \;,
\elabel{Map3-21}
\ee
and make the corresponding replacements in (\ref{eq:Map3-19}), after which one
more angular integration can be carried out, to obtain our final result
\bea
\ave{M^2 M^*}(\theta_1,\theta_2;\theta_3)
&=& S\int{\d y_1\,y_1\over \Theta^2}
\int{\d y_2\,y_2\over \Theta^2} \int_0^{2\pi}{\d\psi\over(2\pi)}
\;\tilde\Gamma^{(3)}_{\rm cen}(y_1,y_2,\psi)
\;{\rm e}^{-Z} \;
\Biggl[ {1\over
24}\,{|\com{q_1}|^{2}|\com{q_2}|^{2}|\com{q_3}|^{2} \over 
\Theta^6} \,f_1^{*2}f_2^{*2} f_3^{2}
\nonumber \\ 
&-&{1\over 9}\,{\com{q_1} \com{q_2} \com{q_3}^{*2}\over \Theta^4}\,
f_1^*f_2^* f_3\, g_3^*
+{1\over 27} \rund{ {\com{q_1}^2 \com{q_2}^2 \com{q_3}^{*4}\over 
|\com{q_1}|^{2}|\com{q_2}|^{2}|\com{q_3}|^{2}
\Theta^2}\, \,g_3^{*2}
+{2\theta_1^2\theta_2^2\over \Theta^4}\,
{\com{q_1} \com{q_2} \com{q_3}^{*2}\over |\com{q_3}|^2\Theta^2} \,f_1^*f_2^*}
\Biggr]\;, 
\elabel{Map3-22}
\eea
which generalizes the result of JBJ for unequal aperture radii. The
proof that this last expression is symmetric with respect to
interchanging $\theta_1$ and $\theta_2$ is the same as the one given
above. 
See Fig.\ \ref{fig3} for an exemplary plot of the latter part of the integrand.

The product of the four $q_i$'s can be written as follows,
\begin{equation}
\com q_1 \com q_2 {\com q_3^{\ast 2}} = \frac 1 {27} \left\{ 2 \left[ y_1^4 +
	y_2^4 + y_1^2 y_2^2 (2 \cos 2\psi - 5) \right] - y_1 y_2 \left[ (y_1^2 +
	y_2^2) \cos \psi +  9 \, \rm{i} (y_1^2 - y_2^2) \sin \psi \right] \right\}
\end{equation}

From the two
complex triple correlators $\ave{M^3}$ and $\ave{M^2 M^*}$, we can now
calculate the four real third-order aperture statistics, in analogy to
what was done in JBJ,
\bea
\ave{M_{\rm ap}^3}(\theta_1,\theta_2,\theta_3)
\!&=&\!\Re\eck{\ave{M^2 M^*}(\theta_1,\theta_2;\theta_3)
+\ave{M^2 M^*}(\theta_1,\theta_3;\theta_2)
+\ave{M^2 M^*}(\theta_2,\theta_3;\theta_1)
+\ave{M^3}(\theta_1,\theta_2,\theta_3)}/4 \;,\nonumber 
\\
\ave{M_{\rm ap}^2 M_\perp}(\theta_1,\theta_2;\theta_3)
\!&=&\!\Im\eck{\ave{M^2 M^*}(\theta_1,\theta_3;\theta_2)
+\ave{M^2 M^*}(\theta_2,\theta_3;\theta_1)
-\ave{M^2 M^*}(\theta_1,\theta_2;\theta_3)
+\ave{M^3}(\theta_1,\theta_2,\theta_3)}/4 \;,\nonumber 
\\
\ave{M_{\rm ap} M_\perp^2}(\theta_1;\theta_2,\theta_3)
\!&=&\!\Re\eck{\ave{M^2 M^*}(\theta_1,\theta_2;\theta_3)
+\ave{M^2 M^*}(\theta_1,\theta_3;\theta_2)
-\ave{M^2 M^*}(\theta_2,\theta_3;\theta_1)
-\ave{M^3}(\theta_1,\theta_2,\theta_3)}/4 \;,
\nonumber 
\\
\ave{M_\perp^3}(\theta_1,\theta_2,\theta_3)
\!&=&\!\Im\eck{\ave{M^2 M^*}(\theta_1,\theta_2;\theta_3)
+\ave{M^2 M^*}(\theta_1,\theta_3;\theta_2)
+\ave{M^2 M^*}(\theta_2,\theta_3;\theta_1)
-\ave{M^3}(\theta_1,\theta_2,\theta_3)}/4 \;, \nonumber\\
\elabel{Map3-23}
\eea
with the same notational convention as used before, e.g., 
$\ave{M_{\rm ap} M_\perp^2}(\theta_1;\theta_2,\theta_3)\equiv
\ave{M_{\rm ap}(\theta_1)M_\perp(\theta_2) M_\perp(\theta_3)}$,
which is symmetric in the last two arguments, as indicated by the
semicolon.  These four expressions have very different physical
interpretations. A significant non-zero value of $\ave{M_{\rm ap}^3}$
indicates that the E-mode of the shear field corresponds to a
convergence field $\kappa$ which has significant skewness. This is the
signal one wants to measure in future cosmic shear surveys, and this
term contains the information about the underlying cosmic density
field, and thus about cosmology. A significant non-zero value of
$\ave{M_{\rm ap} M_\perp^2}$ indicates the presence of a B-mode in the
shear field which is correlated with the E-mode. Although lensing can
generate such a term with small amplitude, by higher-order lensing
effects (caused by source clustering, violation of the Born
approximation in studying light propagation in the Universe, or
multiple light deflections -- see SvWJK for a discussion of these
latter effects), these are probably too small to be
detectable. Therefore, a detection of a $\ave{M_{\rm ap} M_\perp^2}$
most likely will indicate the presence of a `shear' not coming from
lensing, but from, e.g., intrinsic alignment of the galaxies (see,
e.g., Catelan et al.\ 2000; Heavens et al.\ 2000; Crittenden et al.\
2001; Croft \& Metzler 2001; Jing 2002). A significant non-zero value
of $\ave{M_\perp^3}$ indicates that the shear field violates parity
invariance, as a B-mode shear cannot have odd moments if it is
parity-symmetric (Schneider 2003). Finally, a significant non-zero
value of $\ave{M_{\rm ap}^2 M_\perp}$ indicates a parity invariance
violation which is correlated with the E-mode shear field. Neither of
these two latter terms can be explained by cosmic effects which are
expected to by parity-invariant, but either indicates an underestimate
of the statistical errors (coming from the intrinsic ellipticity
distribution of the sources and from cosmic variance), or the presence
of instrumental systematics or artifacts from data reduction (cf.\ the
analogous situation for second-order statistics, where a non-zero
value of $\ave{M_{\rm ap} M_\perp}$ would indicate significant
systematics).

As stated above, measuring the third-order aperture statistics
(through measuring the shear 3PCF and then using the foregoing
relations) yields essentially all the information about the
bispectrum, provided the latter has no sharp features in
$\ell$-space. The analogous statement for the second-order statistics
is not really true: if one considers a cosmic shear survey consisting
of several unrelated fields of size $\Phi$ each, one can calculate the
aperture dispersion from the shear 2PCF for scales, say,
$\theta\lesssim \Phi/4$. However, the cosmic shear field contains
information about the power spectrum of the convergence from all
scales; in particular, due to the fact that the shear 2PCF is obtained
from the power spectrum through a filter function which tends to
constant for $\theta \ell\to 0$, it contains information over the
integrated power on large scales. Hence, the second-order aperture
statistics do not recover the full information about the power
spectrum contained in the shear 2PCF for a survey of a given size. In
order to make better use of the shear data, one should take into
account a shear measure which contains the large-scale power, such as
the top-hat shear dispersion on an angular scale comparable to the
size of the observed fields, say at $\theta=\Phi/4$, which can also be
obtained in terms of the shear 2PCF (CNPT, SvWM). An analogous
situation does not exist for the third-order statistics. This can be
understood intuitively in the following way: Consider again the survey
geometry mentioned above, and assume that to each of the independent
fields a constant shear is added, corresponding to very large-scale
power and/or power in the bispectrum. The aperture measures will be
unable to measure this constant shear, whereas the shear 2PCF will be
sensitive to it, as will be the top-hat shear dispersion. However,
since one cannot form a third-order shear statistics which contains
the shear only, i.e., without reference directions (such as the
direction to the centers of triangles), such a constant shear is
expected to leave no trace on the shear 3PCF. This can be seen
geometrically as follows: consider a triangle of points in a constant
shear field. Rotation of this triangle by 90 degrees changes the sign
of all shear components, and thus the triple product changes sign, for
which reason a constant shear yields no shear 3PCF.  This can also be
seen algebraically from (\ref{eq:Gam0}) and (\ref{eq:Gam1}): The
occurrence of the Bessel functions, which behave like $\ell^6$ and
$\ell^2$, respectively, for small $\ell$ (at fixed $x_i$) removes all
large-scale contributions of the bispectrum in the 3PCF. This fact
suggests that indeed the third-order aperture measures recover
essentially all information about the bispectrum which is present in
the shear field.

One might argue that the skewness of the convergence field, top-hat
weighted in a circular aperture, is sensitive to long wavelength
modes, and so third-order statistics on small scales knows about large
scales. This is true, and may sound like a contradiction to what has
been said above. Looking at the second-order statistics first,
$\ave{\kappa^2}(\theta)$ is sensitive to the power spectrum on all
scales $\ell\lesssim 2\pi/\theta$, and it can be expressed by the 2PCF
$\xi_+$ on angular scales $\le 2\theta$ (CNPT, SvWM). This is due to
the fact that the 2PCF of $\kappa$ is the same as that of
$\xi_+$. Indeed, if one expresses $\ave{\kappa^2}$ in terms of
$\xi_-$, the resulting convolution kernel has infinite support; hence, 
$\ave{\kappa^2}$ cannot be expressed through $\xi_-$ over a finite
range, because $\xi_-$ is {\it not} sensitive to the power spectrum on
large scales.
Something analogous happens for the three-point statistics.
Whereas one can express $\ave{\kappa^3}$ in terms of the shear 3PCF
(this in fact is easily done, e.g. by first expressing $\ave{\kappa^3}$ in
terms of the bispectrum, and then replacing the bispectrum in terms of
the shear 3PCF, using the relations in Sect.\ts\ref{sc:Bispect}), the
integration range is infinite. One cannot calculate $\ave{\kappa^3}$
from the shear over a finite region -- in fact, the mass-sheet
degeneracy does prevent this. Only on very large fields, where the
mean of $\kappa$ can be set to zero, can one in principle measure
$\ave{\kappa^3}$, and that means, one needs information from much
larger scales than the size of the aperture.

\section{\llabel{summary}Summary and Discussion}
In this paper we have considered the relation between the 3PCF of the
cosmic shear and the bispectrum of the underlying convergence
field. Explicit expressions for the (natural components of the) shear
3PCF in terms of the bispectrum have been derived. These expressions
are fairly complicated, and their explicit numerical evaluation
non-trivial. The transformation properties of the 3PCF under parity
reversal can be directly studied using these explicit relations and
confirm those derived in Paper~I by geometrical reasoning.  We have
then inverted these relations, i.e., derived the bispectrum in terms
of the shear 3PCF. Two different expressions were obtained,
corresponding to the two types of natural 3PCF components: one the one
hand $\Gamma^{(0)}$, and $\Gamma^{(i)}$, $i=1,2,3$ on the other
hand. If the shear is due to an underlying convergence field, these
two expressions should yield the same result for the bispectrum; in
general, however, if a B-mode contribution is present, these two
results will differ. Drawing the analogy to the E/B-mode decomposition
for the aperture measures in Sect.~\ref{sc:Gene}, we have conjectured
a linear combination of the two expressions for the bispectrum which
yields the E-mode only. The orthogonal linear combination then yields
the cross-bispectrum of the E-mode with the square of the B-mode
shear. The fact that the bispectrum is real, provided the 3PCF obeys
parity invariance, reaffirms the result of Schneider (2003) that for a
parity-symmetric field, all statistics with an odd power of B-modes
have to vanish.

We have then turned to the aperture statistics, using the filter function that
was suggested by CNPT and also used by JBJ. As a first step we have used the
previously derived expressions for the bispectrum in terms of the 3PCF to
rederive one of the results in JBJ. Then, by considering the third-order
aperture statistics in terms of the underlying bispectrum we have argued that
the third-order aperture statistics with a single filter radius probes the
bispectrum only along a one-dimensional cut through its three-dimensional
range of definition, namely that of equilateral triangles in
$\ell$-space. Generalizing the aperture statistics to three different filter
radii, the full range of the bispectrum can be probed, and, in analogy to JBJ,
we have derived explicit equations for the generalized third-order aperture
statistics in terms of the directly measurable shear 3PCF. We showed
that using different filter radii did not yield additional information
in the case of 
second-order statistics.

The filter function used in the definition of the aperture measures was that
suggested by CNPT. Whereas it does not strictly have finite support, this
disadvantage compared to the filter function defined in SvWJK is outweighed
by the convenient algebraic properties it has; these enabled the explicit
derivation of fairly simple expressions. 

Let us summarize the features of the aperture statistics which render
them so useful as a quantity for characterizing cosmic shear (and other polar
fields):
\bi
\item
The aperture measures can be directly calculated in terms of the shear
correlation functions. It is the latter that can be measured best from
a real cosmic shear survey, as they are not affected by the geometry
of the survey and holes and gaps in the data field. The expressions of
the aperture measures in terms of the shear correlation function are
easy to evaluate by simple sums over the bins for which the
correlation functions have been measured.
\item
The aperture measures provide very localized information about the
underlying power spectrum (in the case of second-order statistics) and
the bispectrum (for third-order statistics) and therefore contain
essentially the full information about the properties of the
underlying convergence field, unless its power in Fourier space has
sharp features (which is not expected for a cosmological mass
distribution, since there is no sharply defined characteristic length
scale).
\item
One can easily calculate the aperture measures in terms of the power
spectrum and the bispectrum, and hence their expected dependence on
the cosmological parameters can be derived and compared to the
measurements. Whereas the aperture measures are just one particular
way to form integral measures of the shear correlation functions -- a
different integral measure was defined by Bernardeau et al.\ (2003)
and applied to a cosmic shear survey in Bernardeau et al.\ (2002) --
it is a particularly convenient one owing to its simple relation to
the bispectrum.
\item
The aperture measures are the easiest way to separate E- and B-modes
of the shear field. Essentially all E/B-mode decompositions for the
second-order shear statistics have been performed using the aperture
measures, and we expect that they will play the same role for the
third-order statistics. Furthermore, since two of the four independent
combinations (\ref{eq:Map3-23}) of the aperture measures are expected
to vanish because of parity invariance, they provide a very convenient way
to detect remaining systematics in the observing, data reduction and
analysis process.
\item 
The aperture statistics are also easily obtained from numerical
ray-tracing simulations, since they are defined in terms of the
underlying convergence in the first place. Hence, in these simulations
one can work directly in terms of the convergence instead of the more
complicated (due to the various components) shear field.
\ei

From the derivation of the 3PCF as a function of the bispectrum, it
becomes clear that the definition of the natural components have eased
the algebra considerably, compared to the case in which one would have
tried to calculate its individual components (like $\gamma_{\rm
tt\times}$). Furthermore, the derivation of the third-order
aperture statistics directly requires the combination of the shear
3PCF provided by the natural components. As discussed in Paper~I there
are various ways to define the natural components of the shear 3PCF,
corresponding to the different centers of a triangle. The derivation
of the aperture measures in terms of the 3PCF has yielded the result
that the projection with respect to the center-of-mass of a triangle
is the most convenient definition (at least in this connection).

\begin{acknowledgement}
We like to thank Mike Jarvis and Lindsay King for helpful comments on
this manuscript. In particular, we are grateful to Mike Jarvis for
pointing out quite a number of typos and errors in the previous
version. We also thank the anonymous referee who, by giving us 
a bit of a hard time, has forced us to make some points
considerably clearer than they were in the original version and 
thus has helped us to improve the paper, not least by asking to
add two figures. 
This work was supported by the German Ministry for
Science and Education (BMBF) through the DLR under the project 50 OR
0106, and by the Deutsche Forschungsgemeinschaft under the project
SCHN 342/3--1.
\end{acknowledgement}

\end{document}